\documentclass[floatfix,superscriptaddress,nofootinbib,aps,twocolumn]{revtex4-1}
\usepackage[dvips]{graphicx}
\usepackage{latexsym,amssymb,amsmath,epsfig,bm,times,psfrag}
\usepackage{color}
\usepackage[bookmarksnumbered,bookmarksopen,colorlinks,citecolor=red,linkcolor=blue]{hyperref}
\usepackage{epstopdf}
\renewcommand\r{\rho}

\renewcommand\o{\omega}
\newcommand\e{\epsilon}
\newcommand\g{\gamma}

\newcommand\m{\mu}
\newcommand\n{\nu}

\newcommand\s{\sigma}
\newcommand\f{\phi}
\newcommand\w{\eta}
\newcommand\ve{\varepsilon}

\renewcommand\L{\Lambda}

\renewcommand\O{\Omega}

\newcommand{\fig}[1]{Fig.~\ref{#1}}
\newcommand{\eq}[1]{Eq.~(\ref{#1})}
\newcommand{\sect}[1]{Sec.~\ref{#1}}

\newcommand\ls{\left[}
\newcommand\rs{\right]}

\newcommand{\lan}{\langle}
\newcommand{\ran}{\rangle}

\newcommand{\non}{\nonumber\\}
\newcommand\pt{\partial}

\newcommand{\diag}{{\rm{diag}}}

\newcommand{\bx}{{\vec x}}
\newcommand{\br}{{\vec r}}
\newcommand{\bp}{{\vec p}}

\newcommand{\bv}{{\vec v}}

\renewcommand{\part}{{\rm part}}
\renewcommand{\vec}{\boldsymbol}
\newcommand{\be}{\begin{equation}}
\newcommand{\ee}{\end{equation}}
\newcommand{\bear}{\begin{eqnarray}}
\newcommand{\eear}{\end{eqnarray}}
\newcommand{\ba}{\begin{array}}
\newcommand{\ea}{\end{array}}

\begin{document}

\title{Thermal vorticity and spin polarization in heavy-ion collisions}

    \author{De-Xian Wei}
    \affiliation{Department of Physics and Center for Field Theory and Particle Physics, Fudan University, Shanghai, 200433, China}
    \author{Wei-Tian Deng}
    \affiliation{School of Physics, Huazhong University of Science and Technology, Wuhan 430074, China}
    \author{Xu-Guang Huang}
    \affiliation{Department of Physics and Center for Field Theory and Particle Physics, Fudan University, Shanghai, 200433, China}
    \affiliation{Key Laboratory of Nuclear Physics and Ion-beam Application (MOE), Fudan University, Shanghai 200433, China}

    \date{\today}

\begin{abstract}
The hot and dense matter generated in heavy-ion collisions contains intricate vortical structure in which the local fluid vorticity can be very large. Such vorticity can polarize the spin of the produced particles. We study the event-by-event generation of the so-called thermal vorticity in Au + Au collisions at energy region $\sqrt{s}=7.7-200$ GeV and calculate its time evolution, spatial distribution, etc., in a multiphase transport (AMPT) model. We then compute the spin polarization of the $\Lambda$ and $\bar{\Lambda}$ hyperons as a function of $\sqrt{s}$, transverse momentum $p_T$, rapidity, and azimuthal angle. Furthermore, we study the harmonic flow of the spin, in a manner analogous to the harmonic flow of the particle number. The measurement of the spin harmonic flow may provide a way to probe the vortical structure in heavy-ion collisions. We also discuss the spin polarization of $\Xi^0$ and $\Omega^-$ hyperons, which may provide further information about the spin polarization mechanism of hadrons.
\end{abstract}

%    \pacs{00.000}
\maketitle

\section{Introduction} \label{sec:intro}
The high-energy heavy-ion collisions provide us the unique opportunity to produce and study the deconfined quark-gluon matter [usually called the quark-gluon plasma (QGP)] in laboratory. Since the first run of the Au + Au collisions at the Relativistic Heavy Ion Collider (RHIC) in 1999, the collected data has revealed a number of striking features of the QGP through measuring a variety of hadronic observables. A set of such observables, considered to constitute the cornerstones of detecting the bulk collective properties of QGP, are the so-called harmonic flow coefficients $v_n$, defined through
\begin{eqnarray}
\label{vn:char}
\frac{d N_{\rm ch}}{d\phi}\propto 1+2v_1\cos{(\phi-\Psi_{1})}+ 2v_2\cos{[2(\phi-\Psi_{2})]}
+\cdots,\non
\end{eqnarray}
where $N_{\rm ch}$ is the number of hadrons of interest (here the charged ones) in a given kinematic (rapidity, transverse momentum $p_T$, etc.) range, $\Psi_{n}$ is the $n$th harmonic plane angle. The harmonic coefficients $v_n$, after carefully subtracting the non-flow contributions, characterize the hydrodynamic response of the final hadronic distribution in momentum space to the spatial shape of the initial-state QGP. The measurement of $v_n$ (especially the second coefficient $v_2$, called elliptic flow coefficient) in non-central collisions reveals that the QGP is a nearly perfect fluid with the lowest ratio of shear viscosity to entropy density ever observed. See recent reviews for more details~\cite{Yan:2017ivm,Romatschke:2017ejr,Shuryak:2014zxa}.

Recently, it was found that such a nearly-perfect fluid is very vortical, namely, the fluid vorticity can be very large. This conclusion was drawn from the measurement of the spin polarization of $\Lambda$ and $\bar{\Lambda}$ hyperons in Au + Au collisions at RHIC~\cite{Abelev:2007zk,STAR:2017ckg,Adam:2018ivw}. The underlying mechanism is the quantum mechanical spin-vorticity coupling, namely, the vorticity can polarize the spin of the constituent particles of the fluid along its direction and therefore the measurement of the spin polarization can deduce the information about the vorticity. Such an idea can be traced back to 2004 although the term vorticity was not mentioned~\cite{Liang:2004ph}; see also Refs.~\cite{Liang:2004xn,Voloshin:2004ha,Gao:2007bc,Huang:2011ru}. The striking experimental finding is that the vorticity averaged over an energy range from $\sqrt{s}=7.7$ GeV to $200$ GeV is of the order of $10^{21} s^{-1}$, surpassing the vorticity of any other known fluid~\cite{STAR:2017ckg}.

The STAR measurement opened the door to a new era of ``subatomic spintronics" where the spin degree of freedom can be used as a probe of the QGP collective property. However, what was measured in Ref.~\cite{STAR:2017ckg} was the spatially averaged vorticity in the mid-rapidity region; the detailed vortical structure was not observed. Only quite recently, the azimuthal structure of the longitudinal and transverse spin polarization of $\L$ and $\bar{\L}$ hyperons at $\sqrt{s}=200$ GeV was reported which provides useful information about the vortical structure~\cite{Adam:2018ivw}. In fact, the vorticity in heavy-ion collisions may receive contributions from different sources which may lead to different vortical structures. One source is the global orbital angular momentum (OAM) of the two colliding nuclei~\cite{Becattini:2007sr,Becattini:2013vja,Csernai:2014ywa,Xie:2015xpa,Xie:2016fjj,Teryaev:2015gxa,Deng:2016gyh,Deng:2016yru,Jiang:2016woz,Ivanov:2017dff,Wang:2017jpl,Shi:2017wpk}. This OAM is perpendicular to the reaction plane if averaged over many events; after the collision, a fraction of the initial OAM is retained in the produced quark-gluon matter in the form of a longitudinal shear flow which results in a finite vorticity. At mid-rapidity region, the vorticity induced by the OAM decreases with increasing beam energy which is consistent with the measured global spin polarization of $\L$ and $\bar{\L}$ hyperons~\cite{Deng:2016gyh,Jiang:2016woz}. The second source of the vorticity is the jet-like fluctuation which could induce smoke-loop type vortex associated with the propagating jet~\cite{Betz:2007kg}. Such generated vorticity is not correlated to the global OAM induced vorticity and thus does not contribute to the global spin polarization of $\L$ and $\bar{\L}$ hyperons. However, on the event-by-event basis, it would contribute to the near-side longitudinal spin-spin correlation~\cite{Pang:2016igs}. The third source of the vorticity is the collective expansion of the fire ball~\cite{Jiang:2016woz,Pang:2016igs,Becattini:2017gcx,Xia:2018tes} which we will discuss in detail in Sec.~\ref{sec:vor}. There may be other sources of the vorticity, for example, the strong magnetic field created by the fast-moving spectators~\cite{Hattori:2016emy,Huang:2015oca} may magnetize the quark gluon matter and generate vorticity through the Einstein-de Haas effect~\cite{Einstein,Huang:2017pqe}.

The purpose of the present paper is two fold. The first one is to give a detailed theoretical study of the so-called thermal vorticity (see definition in \sect{sec:set}) which was shown to be responsible for the spin polarization in an equilibrium plasma~\cite{Becattini:2013fla,Fang:2016vpj}. The second purpose is to study how the vortical structure of the partonic medium can be reflected in the $\L$ and $\bar{\L}$ spin polarization observable when it is represented as a function of the azimuthal angle, transverse momentum, rapidity, etc. In particular, we show that by measuring the harmonic coefficient of the $\Lambda$ polarization in momentum space, one is able to extract important information of the spatial vortical structure of the partonic medium, in a way similar to measuring $v_n$ in Eq.~(\ref{vn:char}) but for spin rather than charge,
\begin{eqnarray}
\label{vn:spin}
\frac{d P}{d\phi}&\propto& 1+2f_1\cos{(\phi-\Phi_{1})}+2f_2\cos{[2(\phi-\Phi_{2})]}+\cdots,\non
\end{eqnarray}
where $P$ denotes the spin polarization which will be defined in Sec.~\ref{sec:set} and $\Phi_n$ is the $n$th harmonic plane for spin. Unlike the electric charge which is a scalar, the spin is a pseudovector so that the above expression can be applied to each component of the spin vector. For the longitudinal component, similar idea has been explored in Ref.~\cite{Becattini:2017gcx}, so we will focus on the transverse components. We note that the harmonic coefficients $f_n$ can be viewed as the spin response to the vortical anisotropy reflecting the collectivity of the spin degree of freedom.

We will also study the spin polarization of $\Xi^0(1314)$ and $\Omega^-(1672)$ baryons. Comparing to $\Lambda$ baryon which contains one valence strange quark, $\Xi^0$ and $\Omega^-$ contain two and three valence strange quarks, respectively. Noticing the fact that the magnetic moments of $\L $, $\Xi^0$, and $\O^-$ are dominated by valence strange quarks, we expect that, among the three baryons, the spin polarization of $\O^-$ can be suppressed the most while $\L$ the least by the magnetic field. Thus the study of spin polarization of $\Xi^0$ and $\O^-$ could be useful for understanding the magnetic-field contribution to spin polarization of hadrons. In addition, the spin of $\O^-$ is $3/2$, so its spin polarization may differ from that of the spin-$1/2$ baryons which also deserves examination.

We note that the fluid vorticity in heavy-ion collisions may induce other novel quantum phenomena which will not be discussed in this paper. Some examples include the polarization of the emitted photons~\cite{Ipp:2007ng}, the vector meson spin alignment~\cite{Liang:2004xn,Abelev:2008ag,Zhou:2017nwi,Singh:2018uad}, the chiral vortical effect~\cite{Kharzeev:2015znc} and chiral vortical waves~\cite{Jiang:2015cva,Chernodub:2015gxa}, the modification to quark-antiquark condensate and phase diagram~\cite{Chen:2015hfc,Jiang:2016wvv,Ebihara:2016fwa,Chernodub:2017ref,Liu:2017spl,Wang:2018sur,McInnes:2018ibt}.

This paper is organized as follows. In Sec.~\ref{sec:set}, we will give a description of the computational method that will be used in the numerical simulation. In Sec.~\ref{sec:vor} and Sec.~\ref{sec:pol}, we will present the main numerical results for the thermal vorticity and the $\L$ and $\bar{\L}$ spin polarization. Finally, we will summarize the main results in Sec.~\ref{sec:dis}. Throughout this paper, we use natural units $\hbar=c=k_B=1$ and the metric $g_{\m\n}=g^{\m\n}=\diag(1,-1,-1,-1)$.

\section{The numerical setup} \label{sec:set}
In non-relativistic hydrodynamics, the fluid vorticity is defined by $\vec\o=(1/2)\vec\nabla\times\bv$ with $\bv$ the flow velocity, which represents the local angular velocity of the fluid cell. The relativistic extension of $\vec\o$ is not unique. One can define different relativistic vorticities according to different physical conditions. A natural one is the kinematic vorticity, $\o^\m=(1/2)\e^{\m\n\r\s} u_\n \pt_\r u_\s$, where $u^\m=\g(1,\bv)$ is the four velocity with $\g=1/\sqrt{1-\bv^2}$ being the Lorentz factor, whose spatial components reduce to the non-relativistic vorticity in low-velocity limit. However, for the purpose of studying spin polarization, it is convenient to use the so-called thermal vorticity tensor~\cite{Becattini:2013fla,Becattini:2015ska},
\begin{eqnarray}
%\label{vort:ther1}
%\varpi^{\m}&=&\frac{1}{2}\e^{\m\n\r\s}u_\n \varpi_{\s\r},\\
\label{vort:ther2}
\varpi_{\m\n}&=&\frac{1}{2}\ls\pt_\n(u_\m/T)-\pt_\m(u_\n/T)\rs,
\end{eqnarray}
where $T$ is the temperature. It was shown that, at local thermal equilibrium, the mean spin vector of spin-$s$ particles (we will consider fermions only so that $2s$ must be an odd integer) with mass $m$ and momentum $p$ produced at point $x$ is given by~\cite{Becattini:2013fla,Fang:2016vpj,Becattini:2016gvu}
\begin{eqnarray}
\label{spin}
S^\mu(x,p)=-\frac{s(s+1)}{6m}(1-n_F)\epsilon^{\mu\nu\rho\sigma} p_\nu \varpi_{\rho\sigma}(x)+O(\varpi)^2,\non
\end{eqnarray}
where $n_F(p_0)$ is the Fermi-Dirac distribution function and $p_0=\sqrt{\bp^2+m^2}$. For $\L$ and $\bar{\L}$ hyperons which are heavy we can take the Boltzmann limit, $1-n_F\approx 1$. In the experiments, the spin of a $\Lambda$ hyperon (similarly for a $\bar{\L}$ hyperon) is measured in its rest frame. Let $S^{*\mu}=(0,{\bm S}^{*})$ denote the spin vector in the rest frame of $\Lambda$. It relates to $S^\mu$ in the laboratory frame by a Lorentz transformation,
\begin{eqnarray}
{\bm S}^{*}={\bm S}-\frac{{\bm p}\cdot{\bm S}}{p_0(p_0+m)}{\bm p}.
\end{eqnarray}
Finally, the spin polarization of $\Lambda$ in the three-direction ${\bm n}$ is defined by
\begin{eqnarray}
\label{def:pol}
P_n= \frac{1}{s}{\bm S}^{*}\cdot{\bm n}.
\end{eqnarray}

In the following sections, we will use the string-melting version of A MultiPhase Transport (AMPT) model to perform the numerical simulations for the thermal vorticity and the spin polarization of $\L$ and $\bar{\L}$ hyperons~\cite{Lin:2004en}. The AMPT model allows us to track each parton's or hadron's position and momentum during the evolution of the system. We use this information to obtain the energy-momentum tensor $T^{\mu\nu}(x)$ by adopting the same smearing function method as that in Ref.~\cite{Deng:2016gyh}. The flow velocity field is defined as the eigenvector of $T^{\mu\nu}$, i.e., $T^{\mu\nu} u_\nu=\varepsilon u_\mu$ ($\varepsilon$ is the energy density), so that in the hydrodynamical term we are using the Landau-Lifshitz frame; see Ref.~\cite{Deng:2016gyh} for more details. The local temperature is extracted from $\ve$~\cite{Lin:2014tya}. Then we use \eq{vort:ther2} to compute the thermal vorticity and use \eq{def:pol} to obtain the spin polarization of the $\L$ and $\bar{\L}$ hyperons. The parameters for the AMPT model are $a=0.55$ and $b=0.15$ GeV$^{-2}$ for the Lund string melting model, strong coupling constant $\alpha_s=0.33$, the Debye screening mass $\mu=2.265$ fm$^{-1}$ which is used in defining the in-medium cross section~\cite{Lin:2014tya,Wei:2018xpm}.

\section{Results for the thermal vorticity} \label{sec:vor}
In this section, we present our numerical results for the thermal vorticity. The results for RHIC Au + Au collisions are obtained by simulating $10^5$ events for each given impact parameter. The coordinates of the colliding system is setup as follows. The $z$ axis is set to be along the beam direction of the projectile, the $x$ axis is along the impact parameter $\vec b$ which points from the target to the projectile, and the $y$ axis is perpendicular to the reaction plane. The origin of the time, $t = 0$, is set to the time when the two colliding nuclei overlap maximally in the beam direction.

\begin{figure}[h]
    \begin{center}
 %   \hspace{-1.5cm}
    \includegraphics[width=5cm]{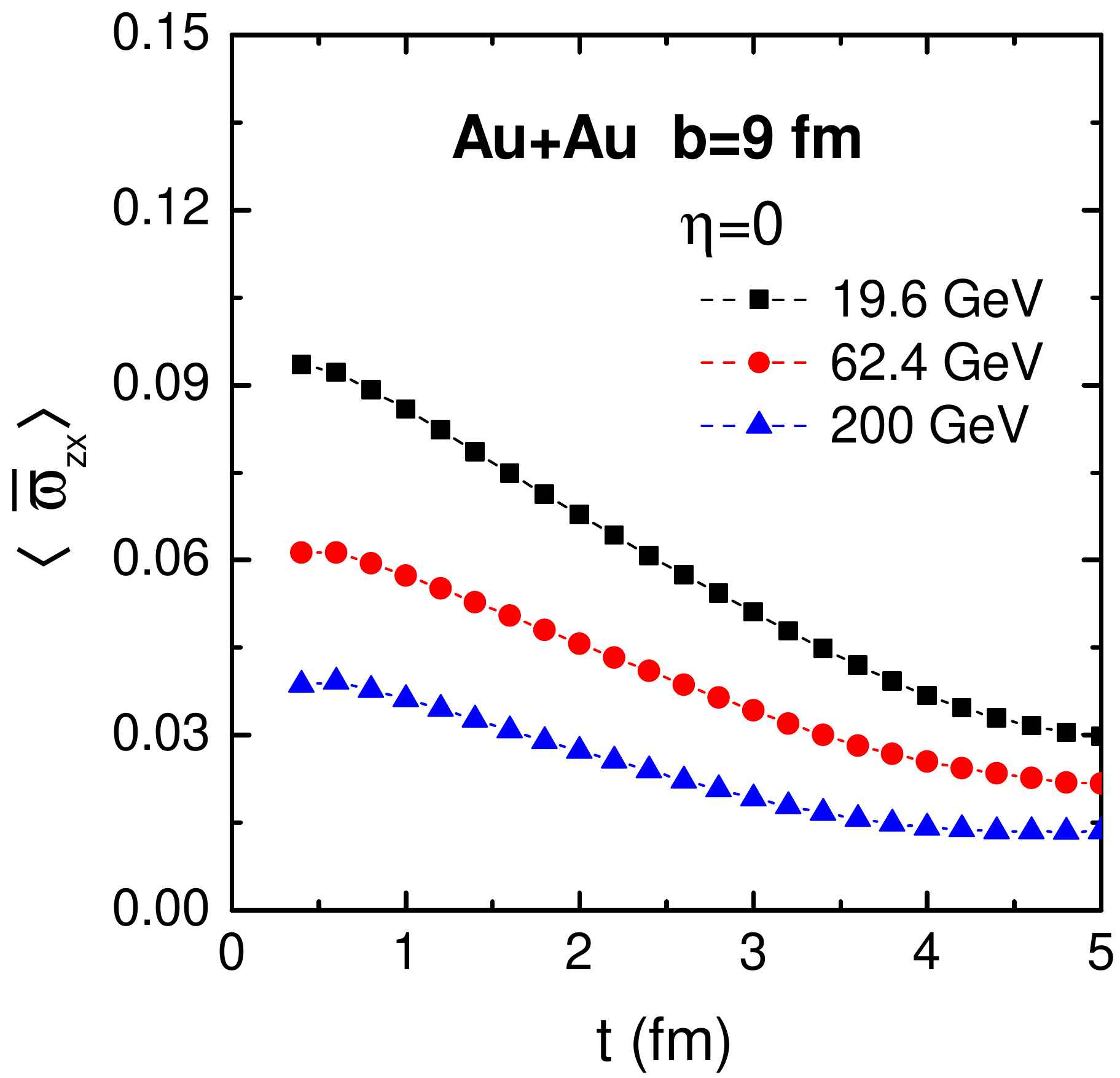} %\\
    \caption{(Color online) The $zx$ component of the thermal vorticity averaged over transverse plane and over colliding events, $\langle\bar\varpi_{zx}\rangle$, as a function of time at $\sqrt{s}=19.6, 62.4$, and $200$ GeV for fixed impact parameter $b=9$ fm and rapidity $\w=0$.}
    \label{tvor:1}
    \end{center}
\end{figure}

The vorticity induced by the global OAM at the collision center in a non-central collision is perpendicular to the reaction plane if averaged over space and events. In \fig{tvor:1}, we show the doubly averaged thermal vorticity $\lan\bar{\varpi}_{zx}\ran$ at mid-rapidity (i.e., $\w=0$ with $\w=\frac{1}{2}\ln[(t+z)/(t-z)]$ the spacetime rapidity) as a function of time for several different collision energies. Here, the double average is defined as the event average of the following energy-density weighted spatial average~\cite{Deng:2016gyh},
\begin{eqnarray}
\bar{\varpi}_{\m\n} = \frac{\int d^2\bx_\perp \ve(\bx_\perp)\varpi_{\m\n}(\bx_\perp)}{\int d^2\bx_\perp \ve(\bx_\perp)}.
\end{eqnarray}

\begin{figure}[h]
%    \begin{center}
%    \hspace{-1.5cm}
    \includegraphics[width=4cm]{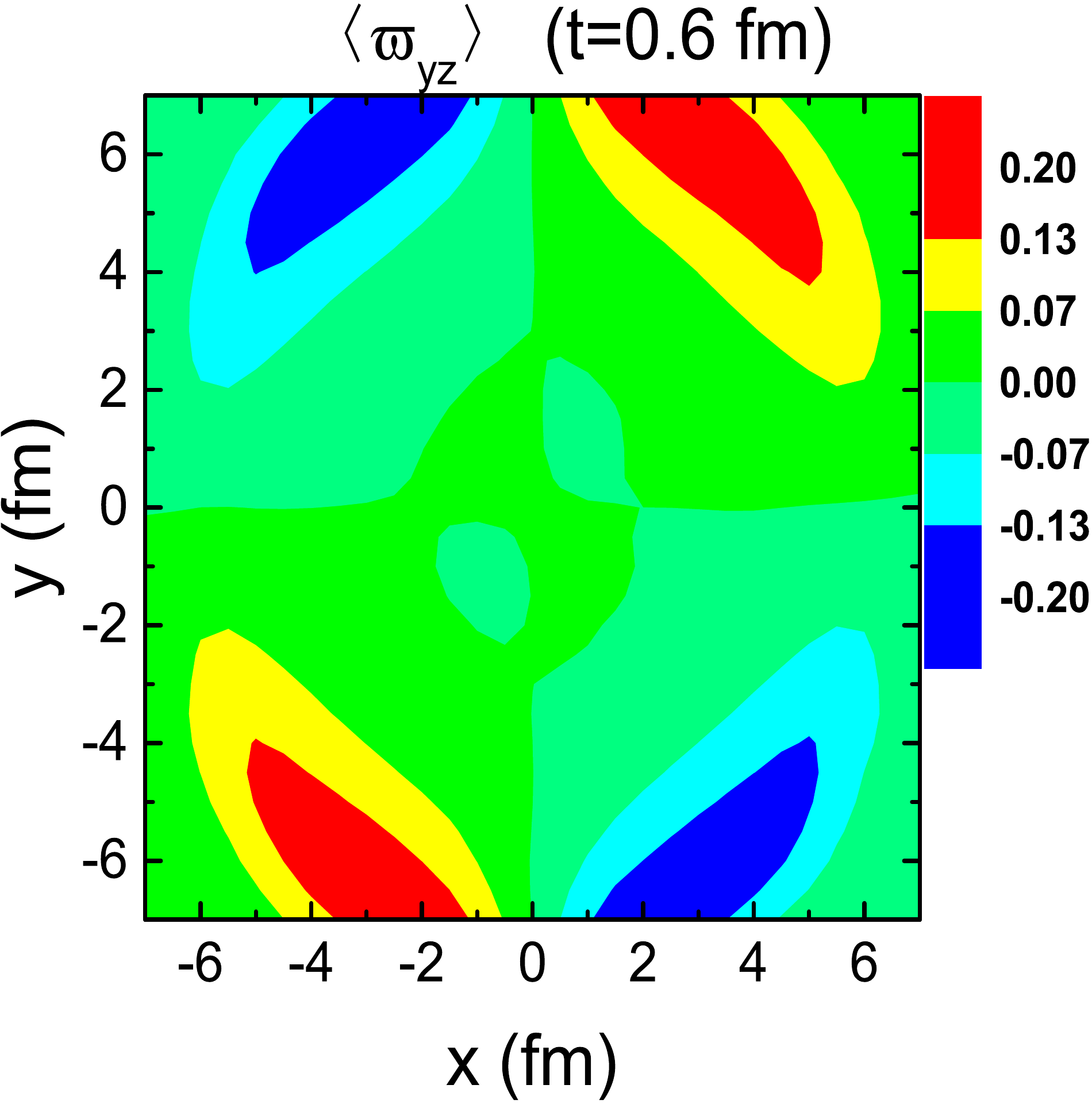} %\\
    \includegraphics[width=4cm]{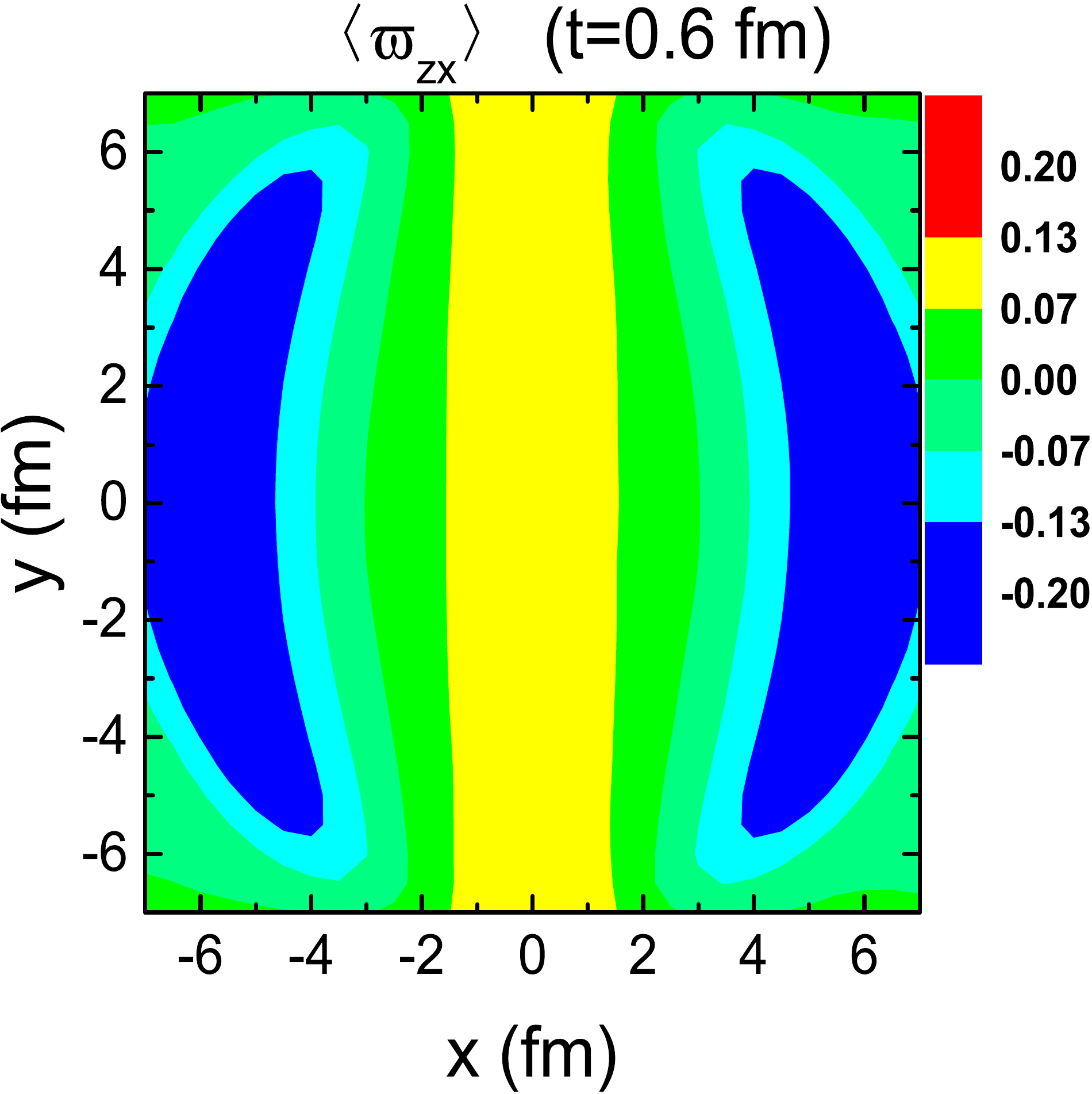}\\ %\\
     \includegraphics[width=4.0cm]{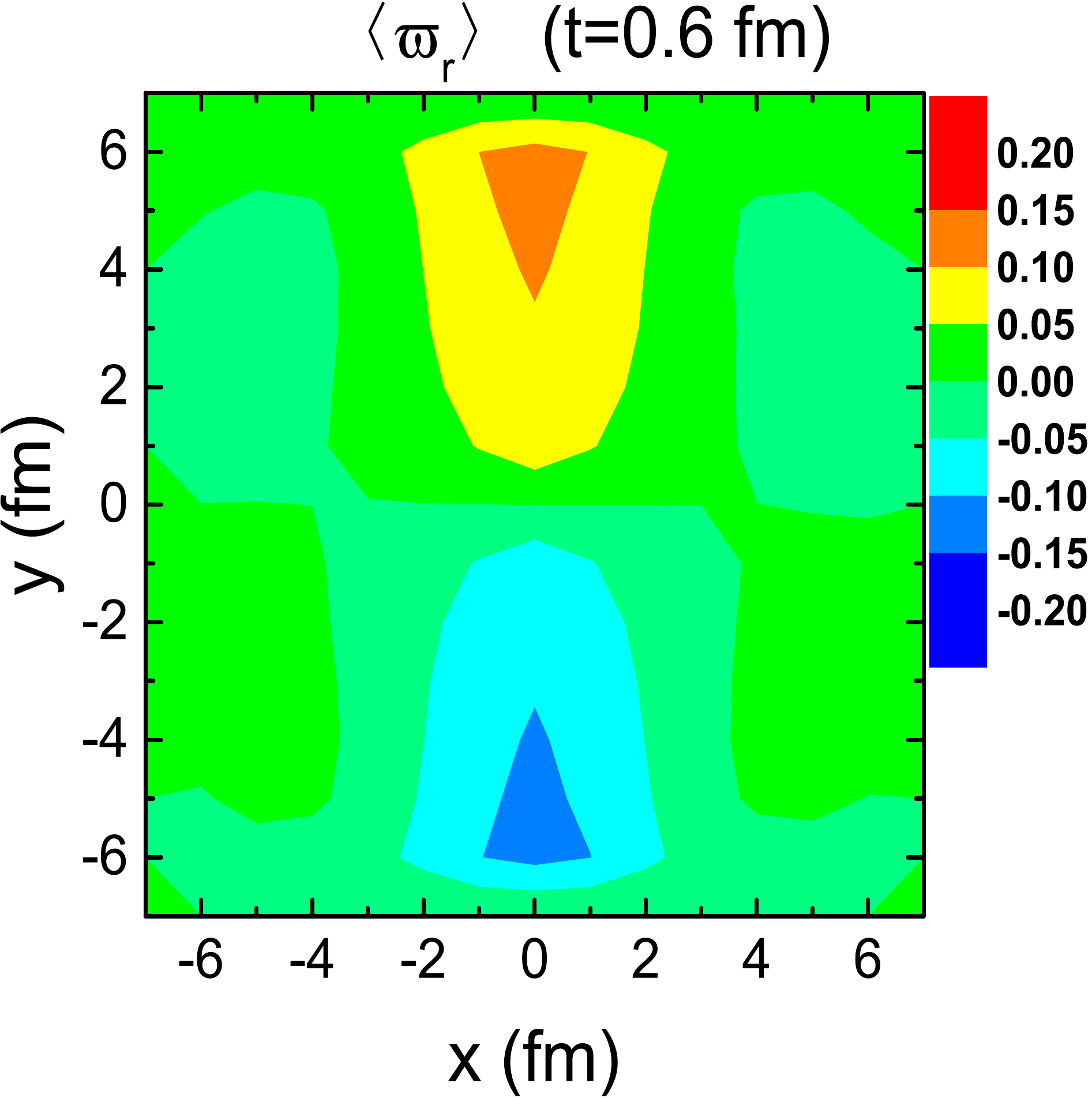} %\\
    \includegraphics[width=4cm]{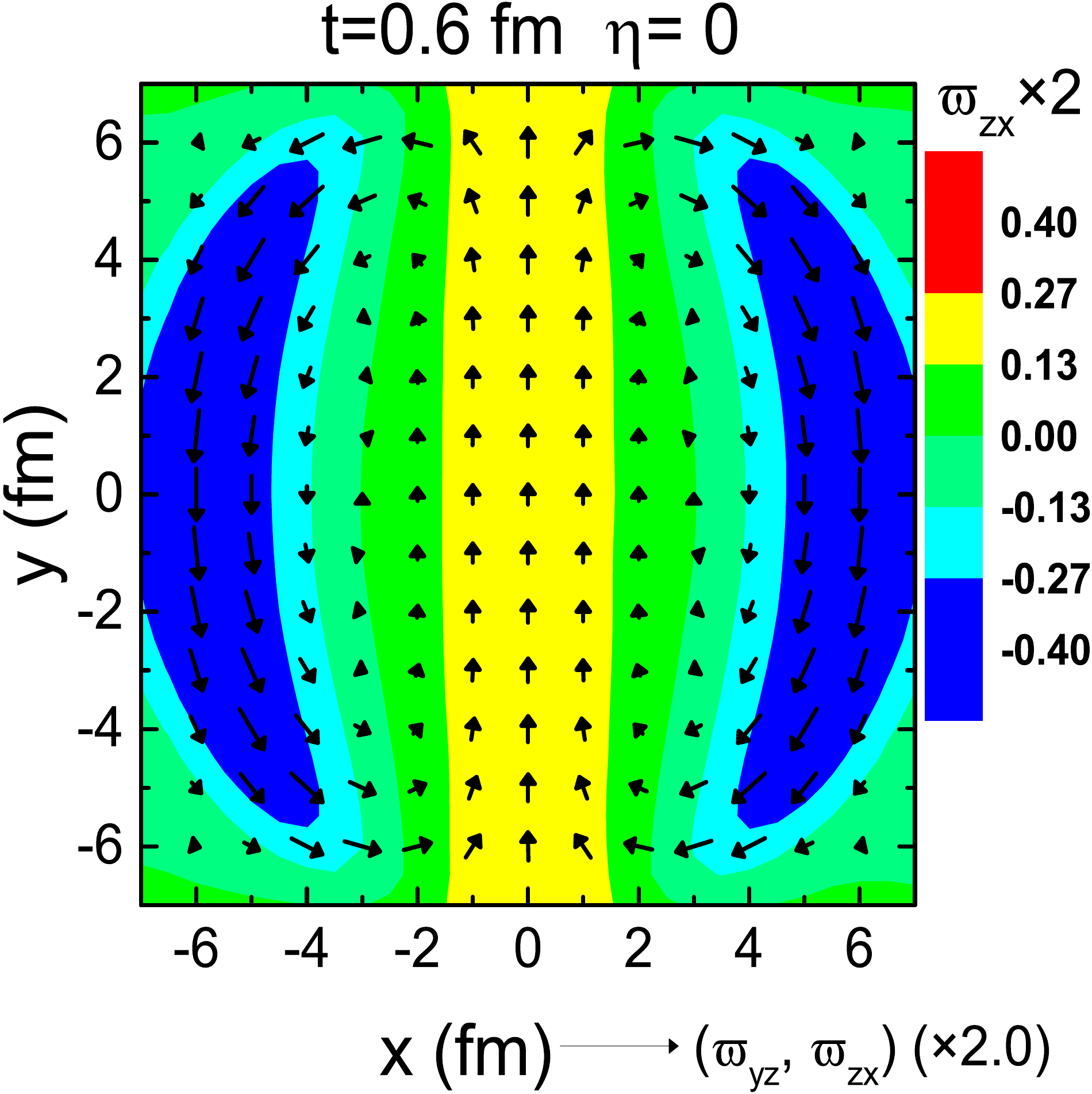}
    \caption{(Color online) The distribution of the $yz$ (top-left panel) and $zx$ (top-right panel) components of the event averaged thermal vorticity in the transverse plane at very early time, $t=0.6$ fm, and rapidity $\w=0$ for Au + Au collisions at $\sqrt{s}=19.6$ GeV averaged over the centrality region 20-50\%. It reflects a vortical structure as displayed by the arrows in the bottom-right panel with the color representing the magnitude $|\vec\varpi_\perp|=(\varpi_{zx}^2+\varpi_{yz}^2)^{1/2}$. The bottom-left panel shows the spatial distribution of the radial thermal vorticity $\hat{\br}\cdot\vec\varpi_\perp$. }
    \label{tvor:ini}
 %   \end{center}
\end{figure}

It is seen that the thermal vorticity at mid-rapidity is smaller at larger collision energy, similar to the kinematic vorticity~\cite{Deng:2016gyh,Jiang:2016woz}. Physically, this can be understood by the fact that at higher collision energy the two colliding nuclei become more transparent to each other leaving the mid-rapidity region closer to the Bjorken boost invariant fluid and thus less vortical. In \fig{tvor:ini}, we plot the event averaged spatial distribution of the $yz$ and $zx$ components of the thermal vorticity at early time, $t=0.6$ fm, and at $\w=0$ in the transverse plane for Au + Au collisions at $\sqrt{s}=19.6$ GeV for centrality region $20-50\%$ (i.e., each figure represents an averaged distribution over the centrality region $20-50\%$). The early-time thermal vorticity is very inhomogeneous in the transverse plane and clear boundary corona effect is seen. Similar structure was also seen in the spatial distribution of kinematic vorticity~\cite{Deng:2016gyh}. Combining the first two panels in \fig{tvor:ini}, we find that the vortex lines in the transverse plane behave like two overlapping, counter oriented, smoke loops; see the bottom-right panel of \fig{tvor:ini} in which we draw the vector plot for $\vec\varpi_\perp=(\varpi_{yz}, \varpi_{zx})$. These two vortex loops are associated with the motion of the participant nucleons in the projectile and target nuclei, respectively. In the bottom-left panel of \fig{tvor:ini} we show the spatial distribution of the radial thermal vorticity $\varpi_r=\hat{\br}\cdot\vec\varpi_\perp$ in the transverse plane. It shows a clear sign separation crossing the reaction plane which may be tested by measuring the $\Lambda$ spin polarization in the radial direction.

\begin{figure}[h]
    \begin{center}
    \includegraphics[width=5.0cm]{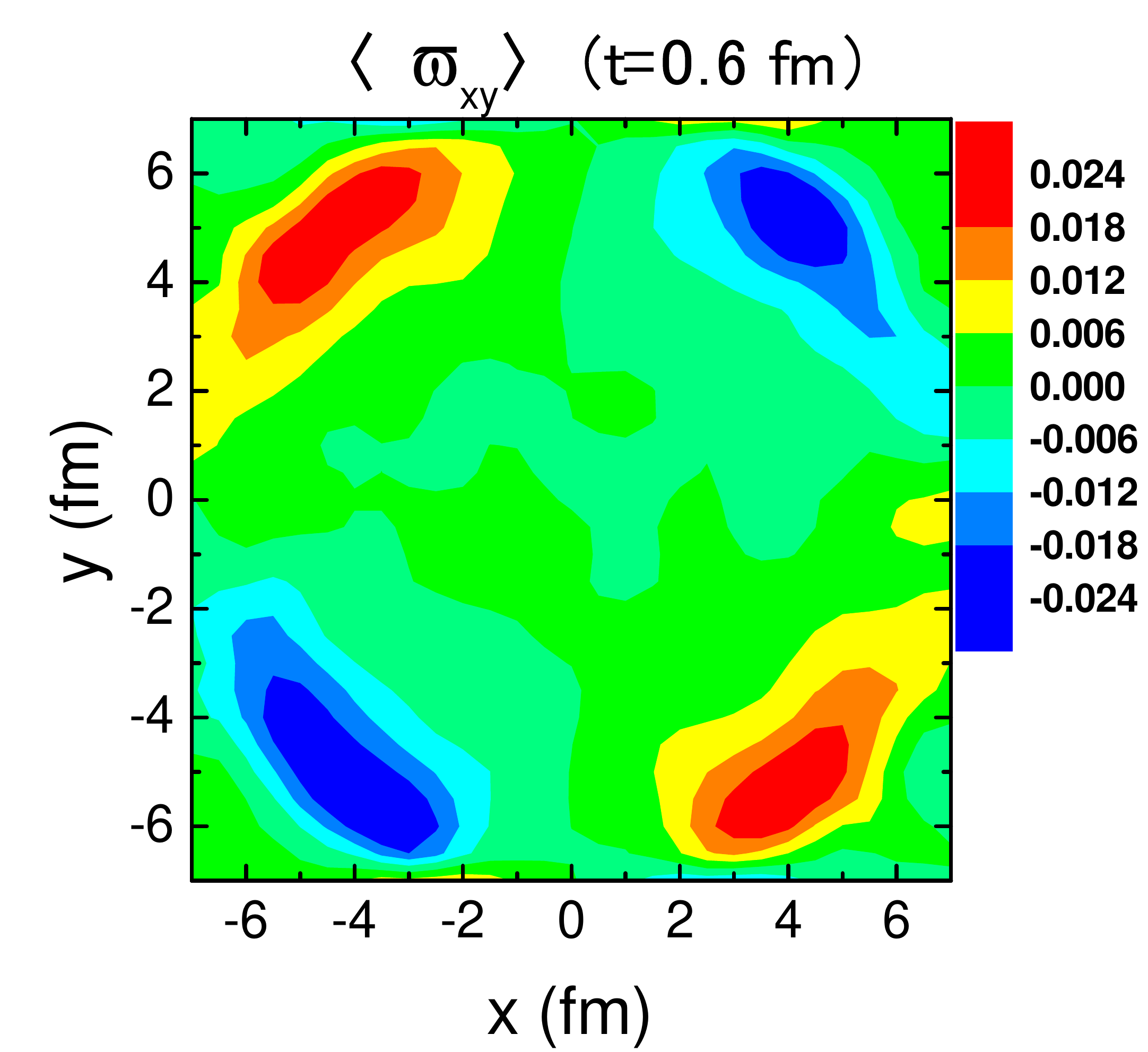} %\\
    \caption{(Color online) The longitudinal component of the event-averaged thermal vorticity distributed in the transverse planeat very early time, $t=0.6$ fm, and rapidity $\w=0$ for Au + Au collisions at $\sqrt{s}=19.6$ GeV averaged over the centrality region 20-50\%. }
    \label{vorlong}
    \end{center}
    \end{figure}
On top of this OAM-induced vorticity, there exist other sources of the vorticity, e.g., the collective expansion of the fire ball, which may induce special patterns in the spatial distribution of the vorticity~\cite{Becattini:2015ska,Jiang:2016woz,Xia:2018tes}. This can be intuitively understood by considering a non-central collision whose velocity profile at a given moment is parameterized by
\begin{eqnarray}
\label{expan}
v_r&\sim&\bar{v}_r(r,z)\left[1+2c_r\cos(2\phi)\right],\nonumber\\
v_z&\sim&\bar{v}_z(r,z)\left[1+2c_z\cos(2\phi)\right],\nonumber\\
v_\phi&\sim&2c_\f \bar{v}_\f(r,z) \sin(2\phi),
\end{eqnarray}
where the reaction plane angle $\Psi_{\rm RP}$ is chosen to be $0$, $r, z, \f$ are the radial, longitudinal, and azimuthal coordinates, and $c_r, c_z$ and $c_\f$ characterize the eccentricity in $v_r, v_z$ and $v_\f$ and are assumed to be constants and small~\footnote{If we consider higher-energy collisions, we can approximately assume a boost invariant longitudinal expansion and a Hubble type transverse expansion at the early stage. It is thus plausible to assume that $\bar{v}_z\propto z/t$, $\bar{v}_r, \bar{v}_\phi\propto rt/R^2$, and $c_z\sim 0$ with $R$ the size of the nucleus~\cite{Deng:2016gyh}. However, for our illustrative purpose, we do not need to make these further assumptions.}.
Subtracting the global OAM effect, the reflection symmetry along the $z$ direction can be assumed which enforces that $\bar{v}_r(r,z)=\bar{v}_r (r,-z)$, $\bar{v}_z(r,z)=-\bar{v}_z (r,-z)$, and $\bar{v}_\f(r,z)=\bar{v}_\f (r,-z)$. Thus, the kinematic vorticity field, $\bm\omega=(1/2)\bm\nabla\times\bm v$, is given by
\begin{eqnarray}
\label{expan:vort}
\omega_r&=&-\ls c_z\frac{2}{r}\bar{v}_z(r,z)+c_\f\frac{\pt \bar{v}_\f(r,z)}{\pt z}\rs \sin(2\phi),\nonumber\\
\omega_z&=&\ls c_r\frac{2}{r}\bar{v}_r(r,z) + \frac{c_\f \bar{v}_\f (r,z)}{r} + c_\f\frac{\pt \bar{v}_\f (r,z)}{\pt r}\rs\sin(2\phi),\nonumber\\
\omega_\phi&=&\frac{1}{2}\frac{\partial\bar{v}_r (r,z)}{\partial z}[1+2c_r\cos(2\phi)]  \nonumber \\
 && -\frac{1}{2}\frac{\partial\bar{v}_z (r,z)}{\partial r}[1+2c_z\cos(2\phi)].
\end{eqnarray}
Thus we find that if we subtract the global OAM contribution, at mid-rapidity, $\w=0$, only the longitudinal vorticity $\o_z$ can be nonzero while the transverse components $\omega_r$ and $\o_\f$ vanish. At finite rapidity, all three components of $\vec\o$ can be finite and the transverse vorticity is dominated by the $\f$ component. Note that $\o_r$ and $\o_z$ show quadrupole structures in the transverse plane. The longitudinal vorticity at $\w=0$ is depicted in \fig{vorlong} for the centrality region 20-50\% for Au + Au collisions at $\sqrt{s}=19.6$ GeV in which a clear quadrupolar structure is seen. Such longitudinal vorticity at the mid-rapidity region has been carefully examined and it was found that it can induce a sizable longitudinal spin polarization of $\L$ and $\bar{\L}$ hyperons even at high $\sqrt{s}$ where the transverse vorticity is expected to be small~\cite{Becattini:2017gcx}; see also Refs.~\cite{Xie:2016fjj,Xia:2018tes}. The azimuthal vorticity, $\o_\f$, is special because it can be finite even for central collisions and we will study it in more detail.
%It can induce a vortical quadrupole in the reaction plane at $\w\neq 0$ as shown in \fig{vorqua}.

\begin{figure}[h]
    \begin{center}
%    \hspace{-1.0cm}
    \includegraphics[width=4.0cm]{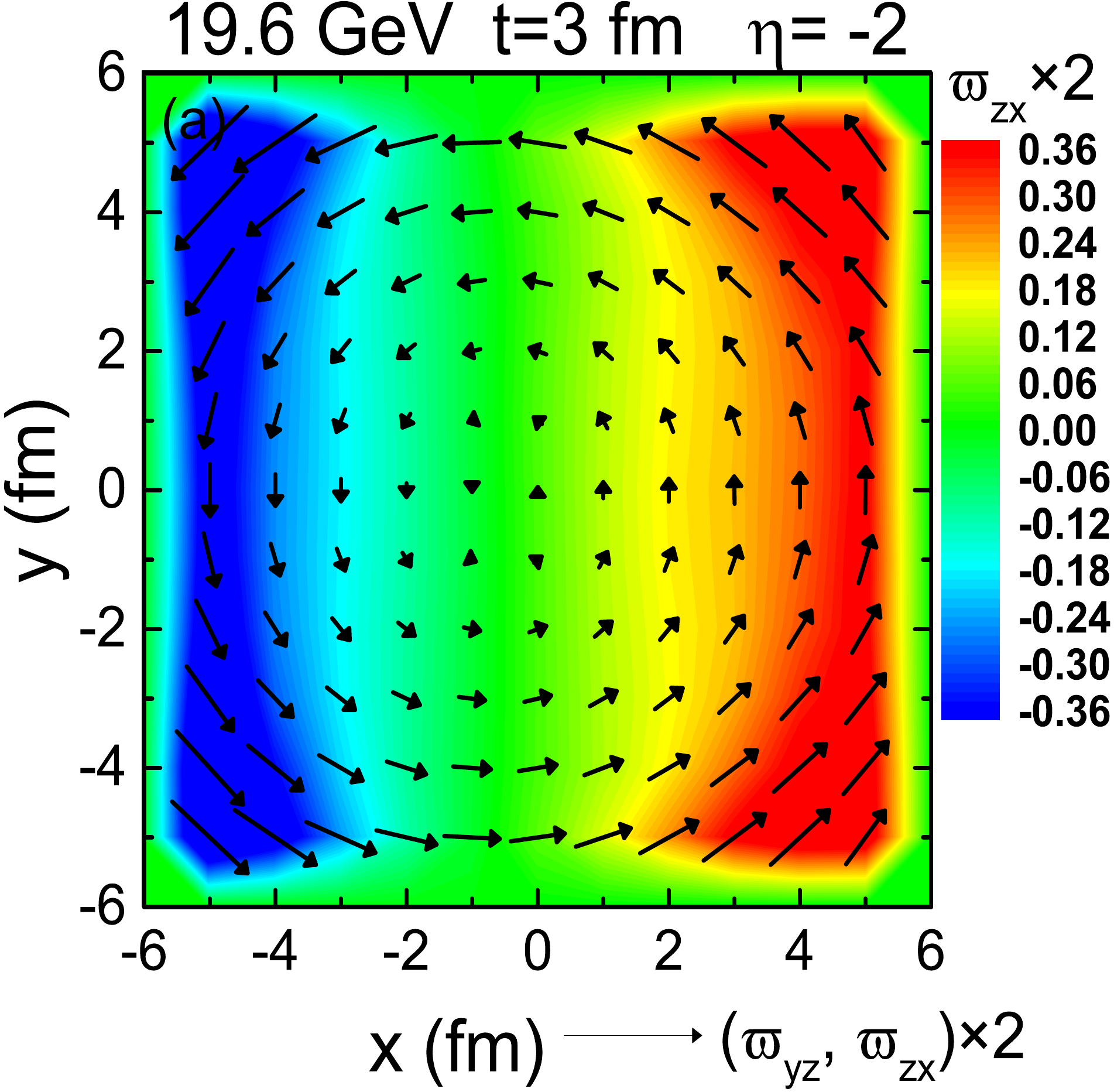} %\\
    \hspace{0.0cm}
    \includegraphics[width=4.0cm]{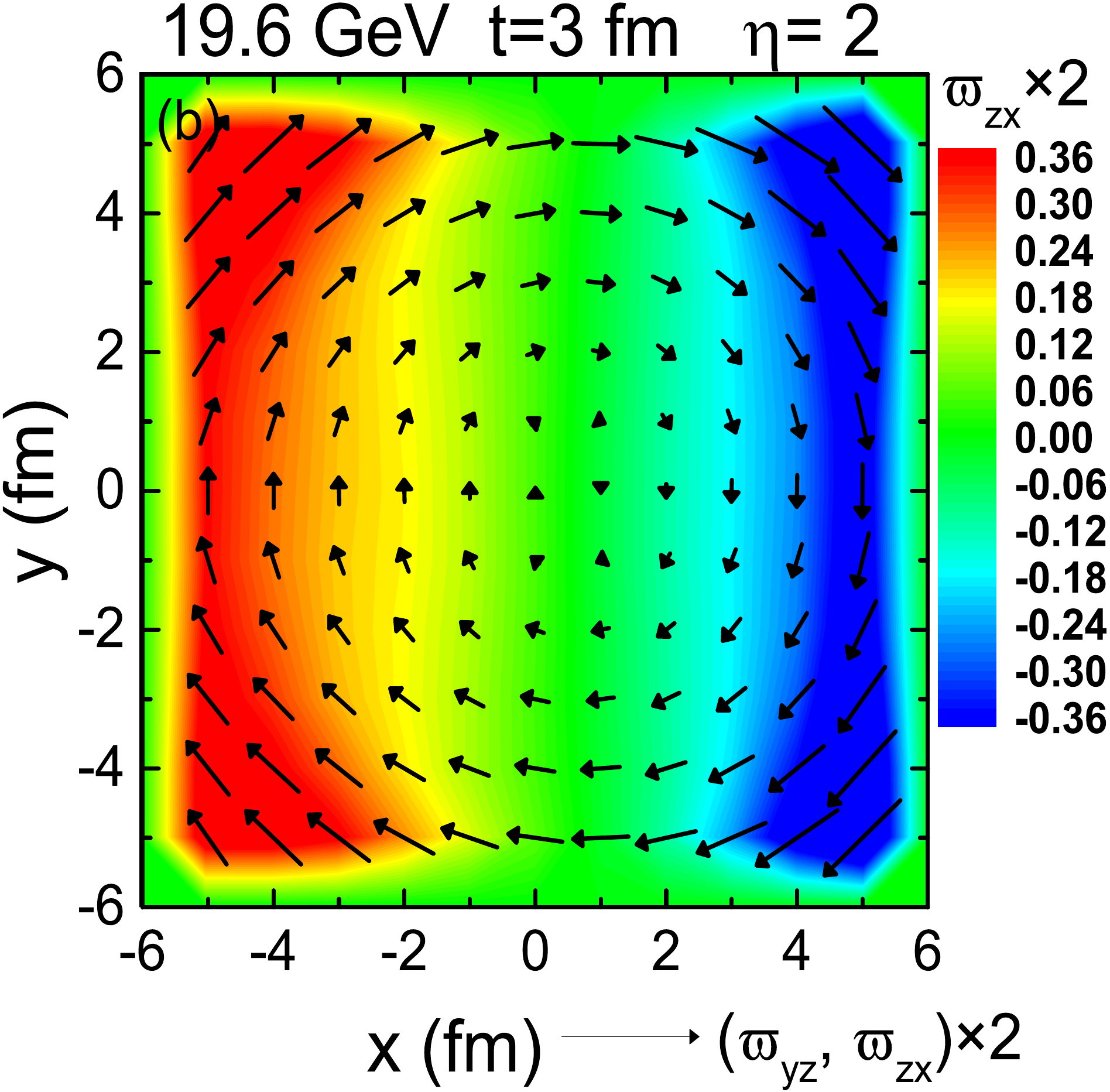} %\\
    \hspace{0.0cm}
    \includegraphics[width=4.0cm]{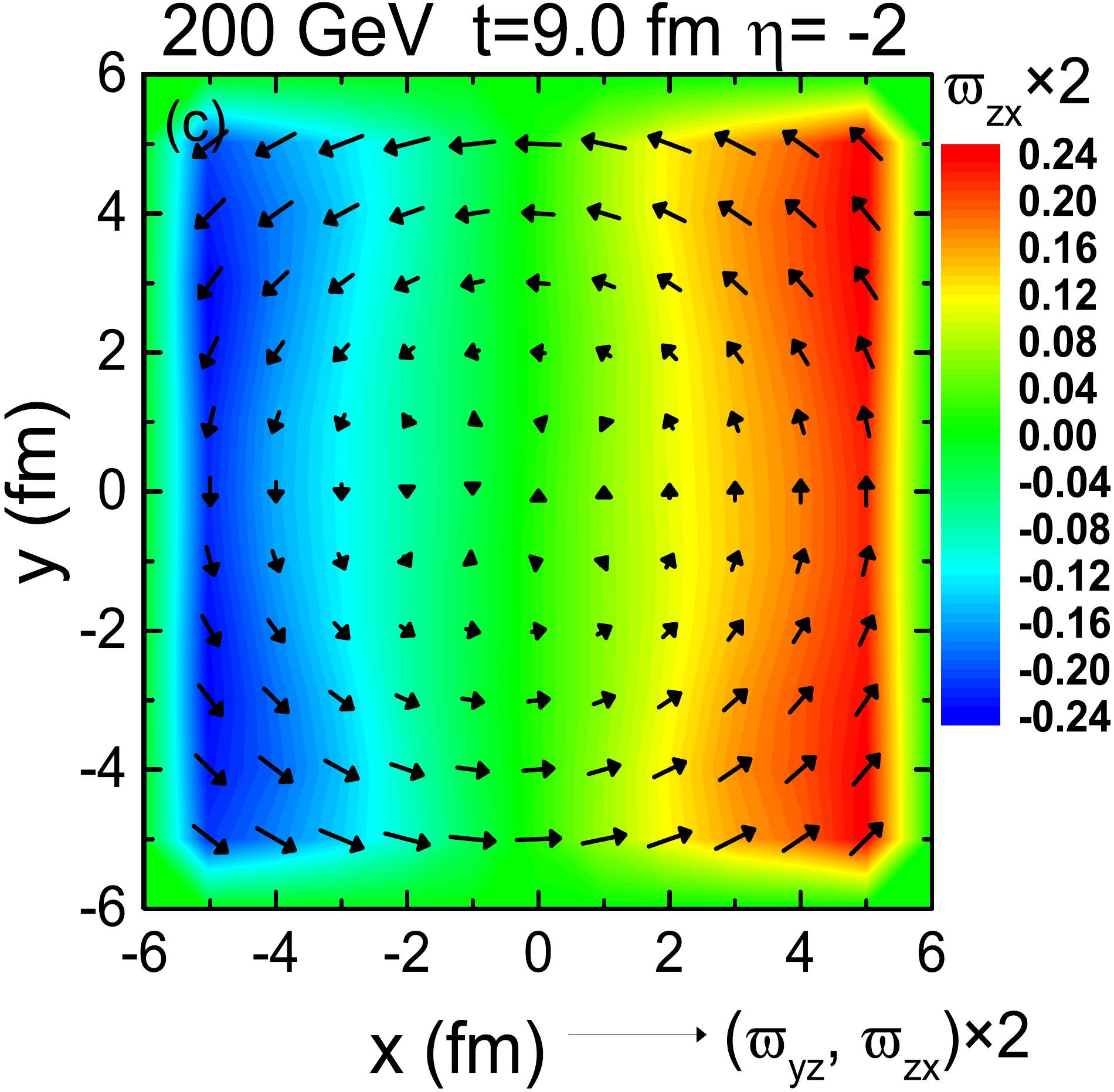} %\\
    \hspace{0.0cm}
    \includegraphics[width=4.0cm]{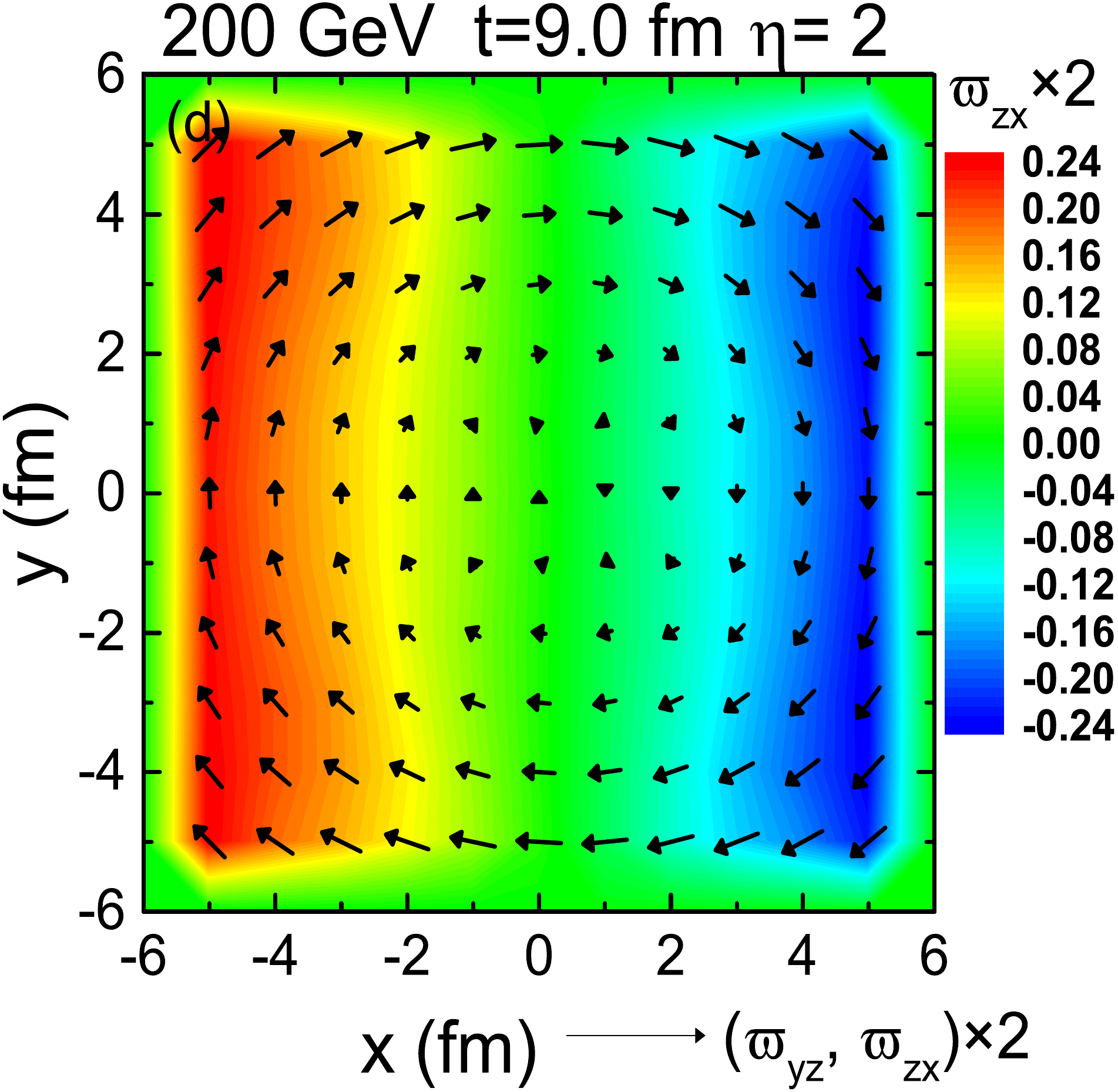} %\\
    \caption{(Color online) The vector plot for the thermal vorticity projected to the transverse plane at spacetime rapidity $|\eta|=2$ for Au + Au collisions at 19.6 and 200 GeV averaged over events in 20-50\% centrality range at fixed time $t=3$ and $9$ fm, respectively. The background color represents the distribution of the $\varpi_{zx}$ component.}
    \label{vorloop}
    \end{center}
    \end{figure}
\begin{figure}[h]
    \begin{center}
%    \hspace{-1.5cm}
    \includegraphics[width=4.0cm]{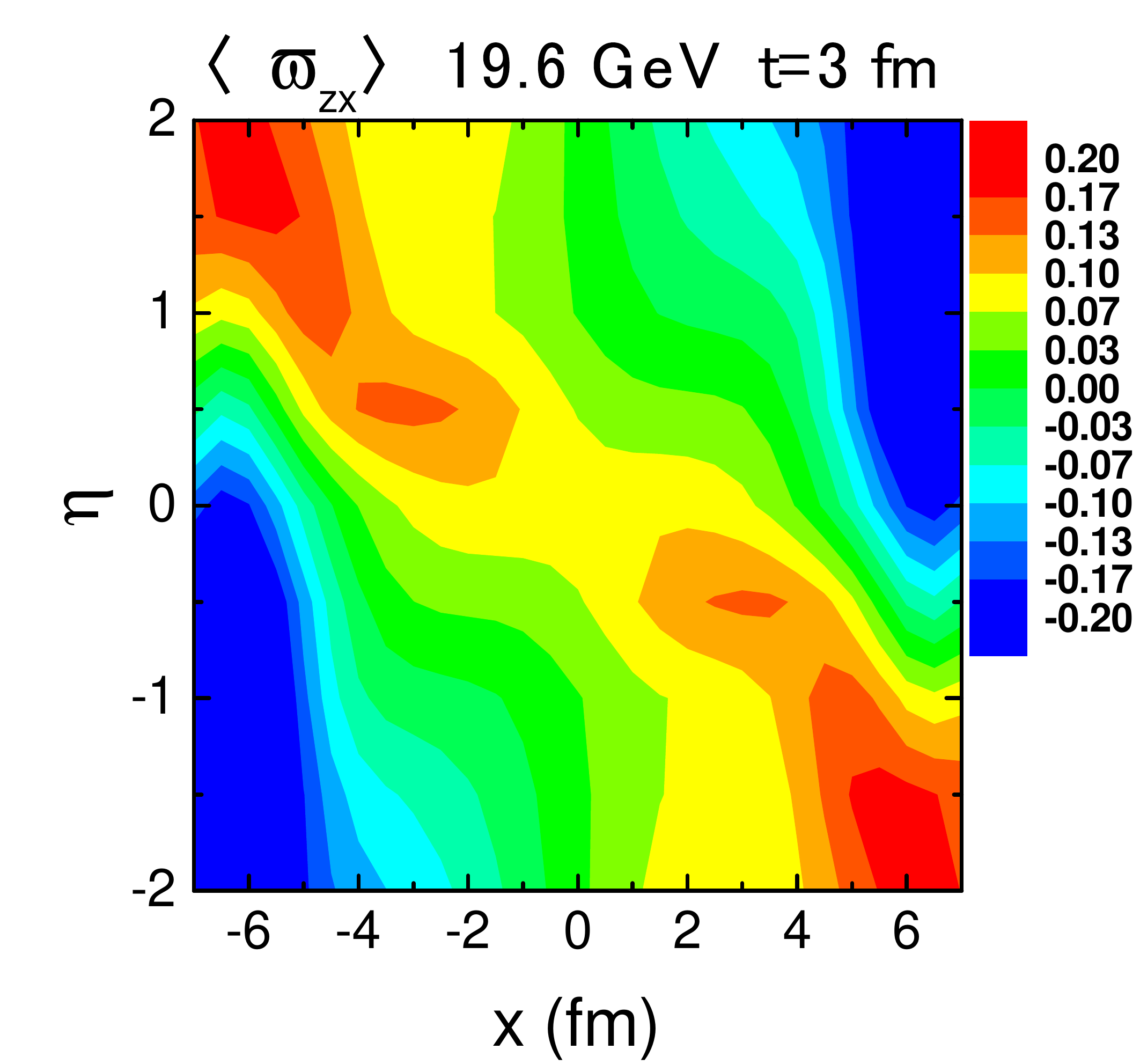} %\\
    \includegraphics[width=4.0cm]{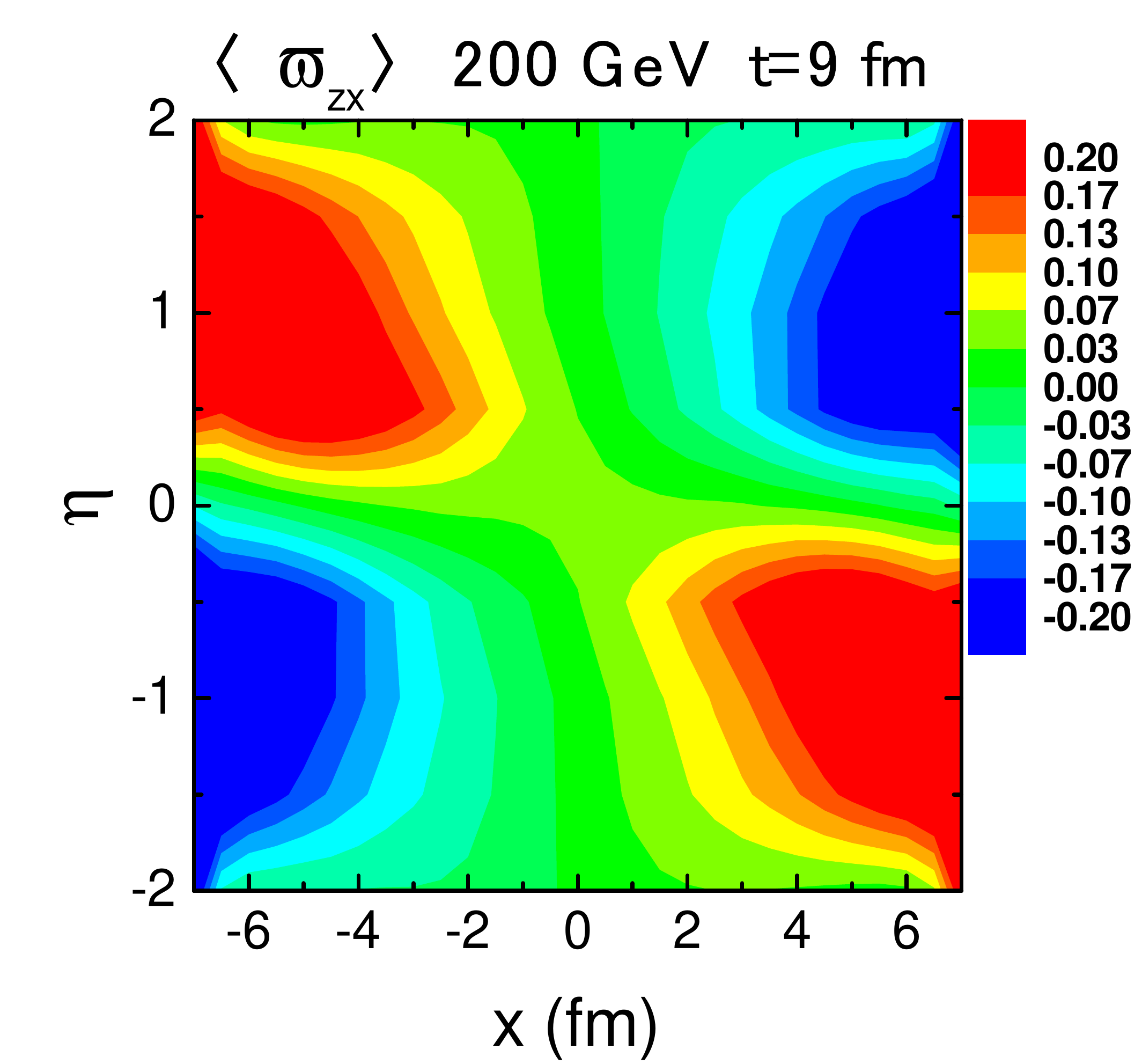}
    \caption{(Color online) The distribution of event-averaged thermal vorticity in the $x$-$\w$ plane for Au+Au collisions at 19.6 and 200 GeV, respectively. Other parameters are the same as \fig{vorloop}.}
    \label{vorqua}
    \end{center}
\end{figure}

Although the above illustrative discussion is for the kinematic vorticity, we expect that similar azimuthal and spacetime rapidity dependence hold also for the thermal vorticity. In \fig{vorloop}, we show our numerical simulation of the transverse thermal vorticity at $\w=2$ and $\w=-2$ at late time $t=3$ and $9$ fm for Au + Au collisions at $\sqrt{s}=19.6$ and $200$ GeV, respectively. The arrows represent the vortex lines projected to the transverse plane and the background color represents the component $\varpi_{zx}$. Clearly seen is the smoke-loop type vortex lines oriented in opposite directions for opposite spacetime rapidity; similar pattern was obtained also in Refs.~\cite{Xie:2017upb,Xia:2018tes}. Such circular vortical structure at finite rapidity gives a quadrupolar distribution of the $\varpi_{zx}$ component in the $x$-$\w$ plane as shown in \fig{vorqua}. Similar results were also discussed in Refs.~\cite{Csernai:2014ywa,Becattini:2015ska,Teryaev:2015gxa,Jiang:2016woz,Li:2017slc,Shi:2017wpk,Kolomeitsev:2018svb,Xia:2018tes}.

\section{Spin polarization of hyperons} \label{sec:pol}
\begin{figure}[h]
    \begin{center}
 %   \hspace{-1.5cm}
    \includegraphics[width=4cm]{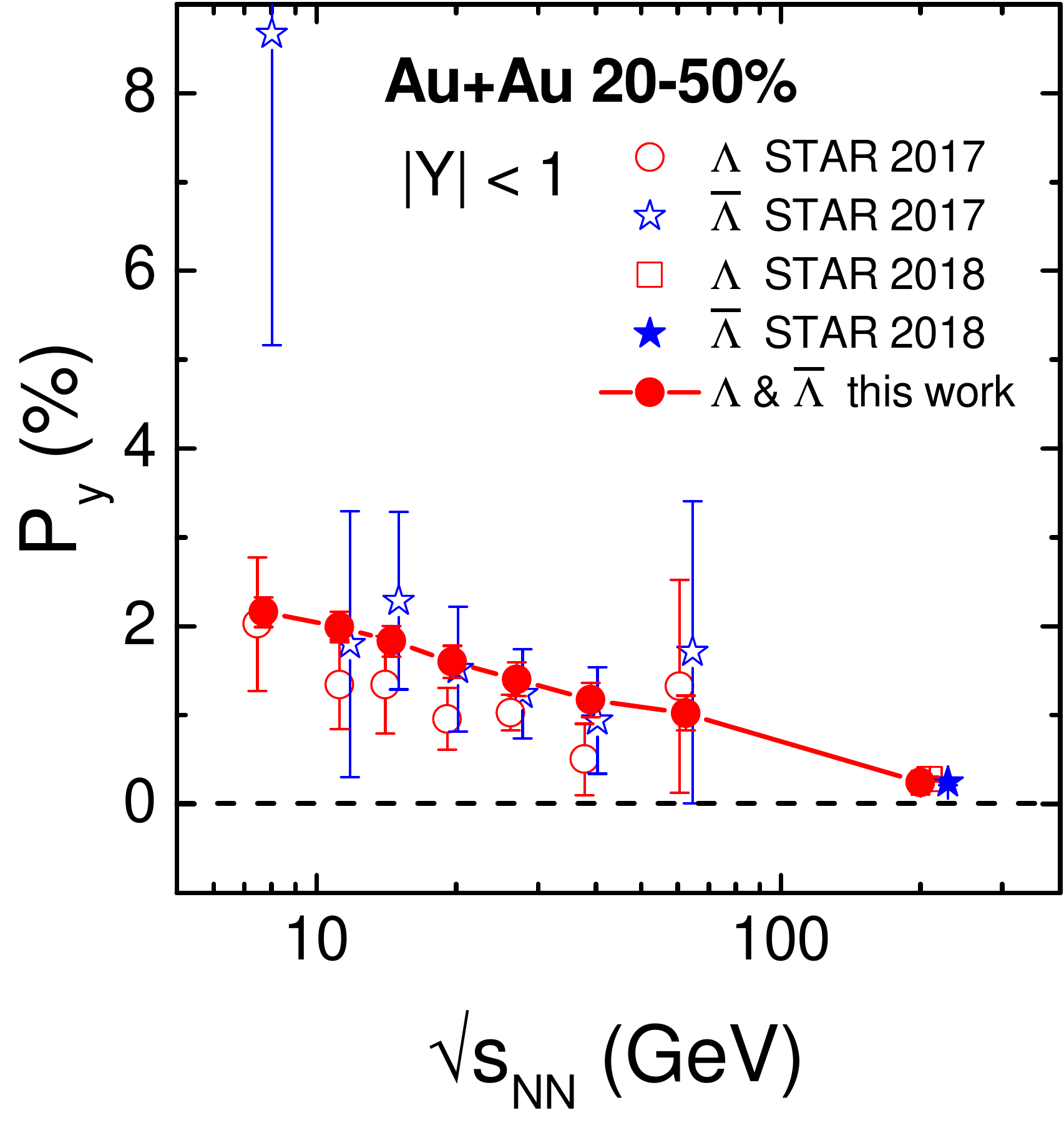} %\\
    \includegraphics[width=4cm]{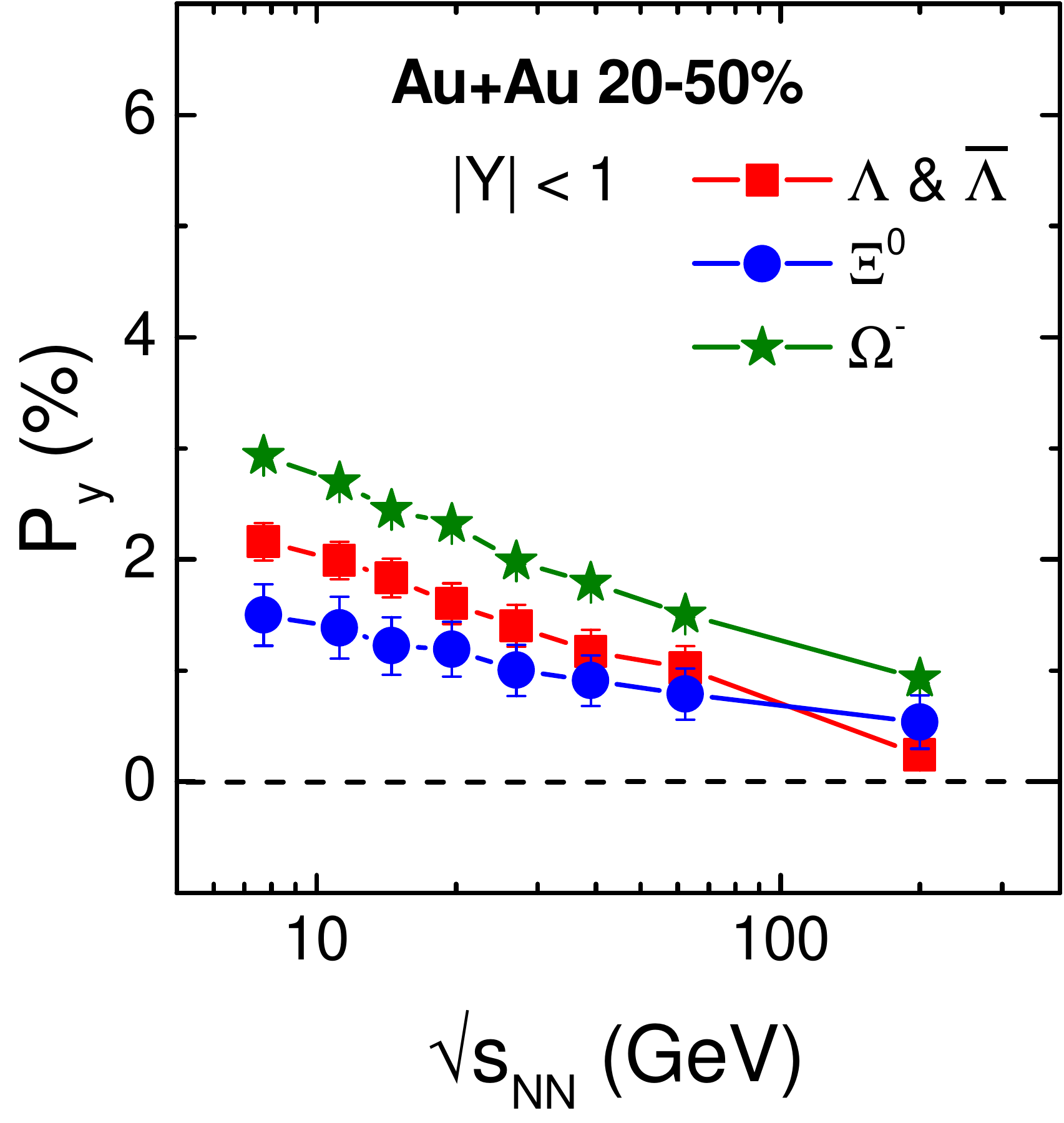} %\\
    \caption{(Color online) (Left) The averaged $\L$ and $\bar{\L}$ spin polarization along $y$ direction in 20-50\% centrality range of Au+Au collisions as a function of collision energy. The rapidity window for $\L$ and $\bar{\L}$ is $|Y|<1$. Open points: STAR data~\cite{STAR:2017ckg,Adam:2018ivw}. Red solid points: this work. (Right) The spin polarization $P_y$ for $\Xi^0$ and $\O^-$. Other parameters are the same as the left panel. }
    \label{globpol}
    \end{center}
\end{figure}

\begin{figure}[h]
    \begin{center}
%    \hspace{-1.5cm}
%   \;\;\;\;\;\;\;\; \includegraphics[width=4cm]{fig107a} %\\
%    \;\;\;\;\includegraphics[width=4cm]{fig107b} %\\
    \includegraphics[width=4cm]{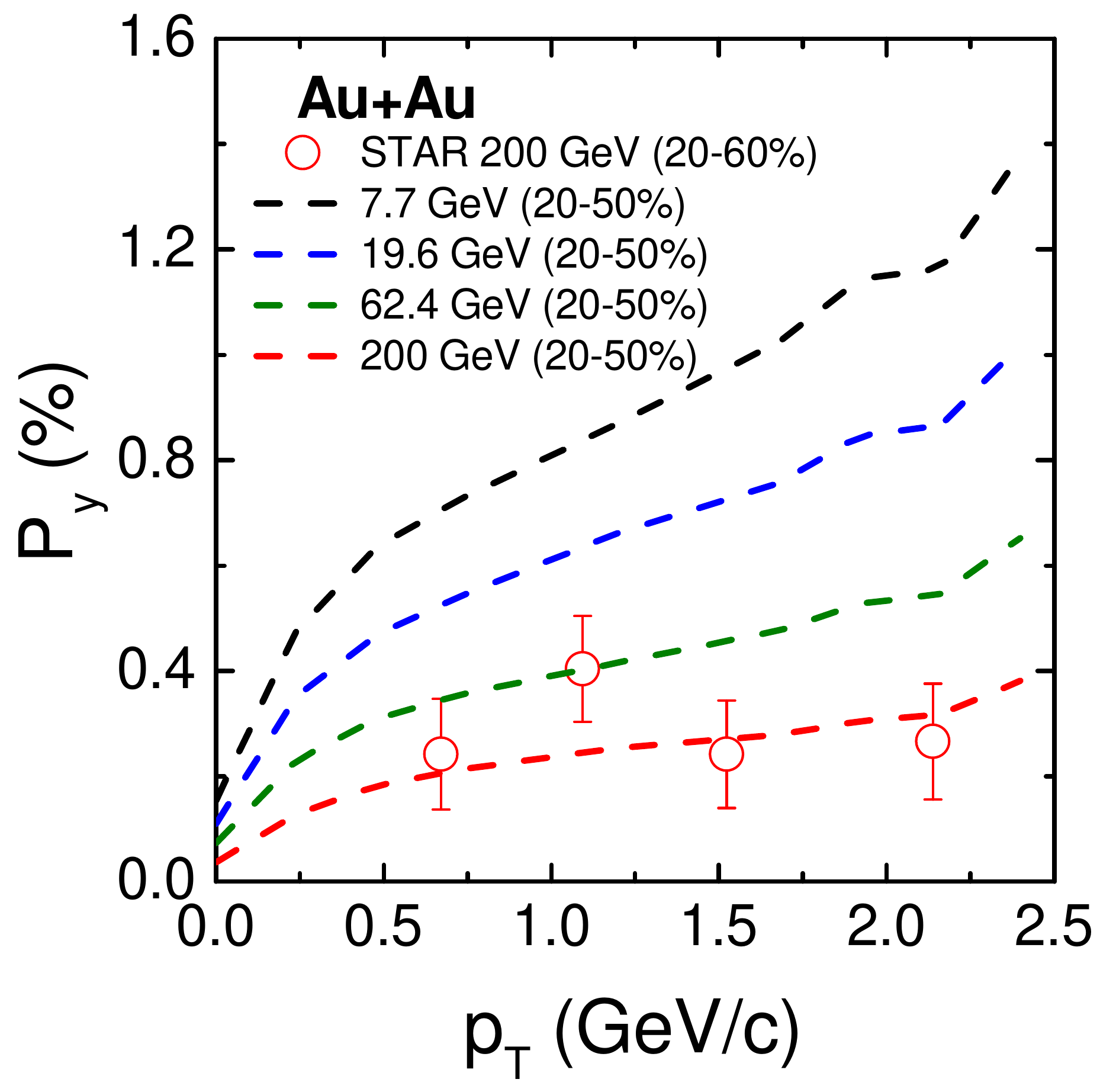} %\\
    \includegraphics[width=4cm]{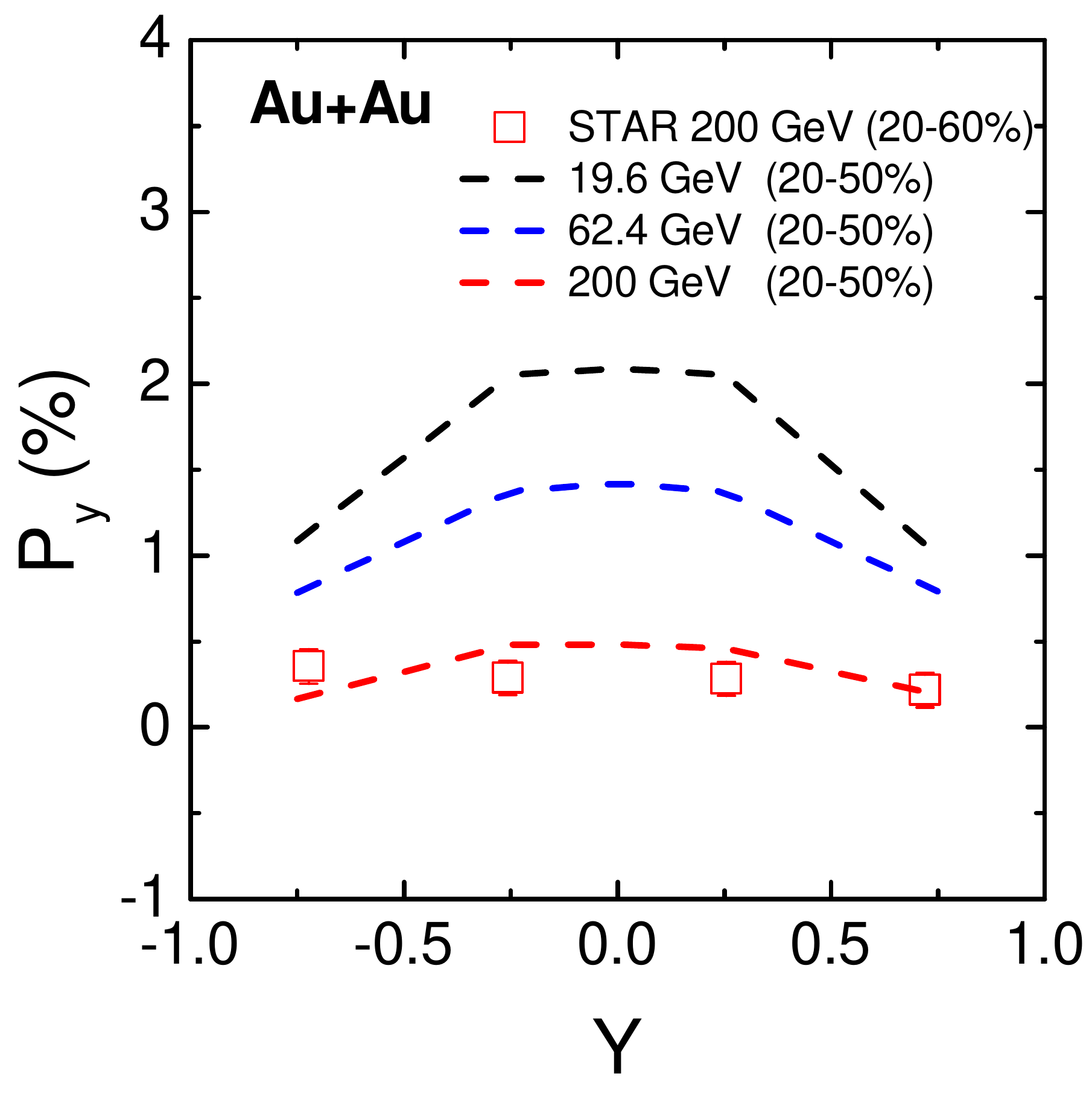} %\\
    \caption{(Color online) The $p_{T}$ and rapidity dependence of the global polarization at different collision energies. Open points: STAR data~\cite{Adam:2018ivw}. Dotted lines: this work.}
    \label{ptrap}
    \end{center}
\end{figure}

The spatial structure of the thermal vorticity discussed in \sect{sec:vor} can be transformed into the structure of the spin polarization of $\L$ and $\bar{\L}$ hyperons in momentum space. In \fig{globpol} (left) we show our result for the global spin polarization of $\L$ and $\bar{\L}$ hyperons along the $y$ direction, i.e., the direction of the total OAM, for Au + Au collisions in the centrality region 20-50\% and rapidity region $-1<Y<1$ from $\sqrt{s}=7.7$ to 200 GeV, where $Y=\frac{1}{2}\ln[(p_0+p_z)/(p_0-p_z)]$. Within the error bars, our numerical result is consistent with the experimental data except for $7.7$ GeV where the data for ${\bar{\L}}$ is very large. We do not take into account the possible feed-down contributions to the global polarization; the previous estimate showed that including such contributions will suppress the $\L$ and $\bar{\L}$ polarization by about $10-20$\%~\cite{STAR:2017ckg,Becattini:2016gvu,Karpenko:2016jyx,Li:2017slc,Kolomeitsev:2018svb}. Comparing to \fig{tvor:1}, we emphasize that the energy dependence of $P_y$ is consistent with that of $\varpi_{zx}$. We also depict the $p_T$ and rapidity $Y$ dependence of the global polarization and compare to the experimental data in \fig{ptrap}. The results show different patterns as those simulated in Ref.~\cite{Sun:2017xhx}. The rapidity dependence is qualitatively consistent with the spacetime-rapidity dependence of fluid vorticity~\cite{Deng:2016gyh}. Within error bars, consistence between the data~\cite{Adam:2018ivw} and our simulation is seen.

In \fig{globpol} (right) we draw the spin polarization of $\Xi^0$ and $\O^-$ for Au+Au collisions in 20 - 50\% centrality range and rapidity window $|Y|<1$ . The results are similar with that of $\L$ and $\bar{\L}$ and can be understood by noticing the mass ordering and spin ordering among $\L$, $\Xi^0$, and $\O^-$: $m_{\L}< m_{\Xi^0} <m_{\O^-}$ and ${\rm spin}(\O^-)=3/2$, ${\rm spin}(\Xi^0)={\rm spin}(\L)=1/2$.  According to \eq{spin} and \eq{def:pol}, lighter and higher-spin particles are easier to be polarized by the fluid vorticity. The study of $\Xi^0$ and $\O^-$ polarization may also provide useful information for the understanding of the magnetic field contribution to the spin polarization of hadrons. This is because that the valence quark contents of $\L$, $\Xi^0$, and $\O^-$ are $uds$, $uss$, and $sss$, respectively, and their magnetic moments are all dominated by strange quarks, $\m_\L\approx \m_s$, $\m_{\Xi^0}\approx 2
\m_s$, and $\m_{\O^-}\approx 3\m_s$. As $\m_s\approx -0.613 \m_N< 0$, the magnetic field (which is roughly along the same direction as the OAM) will give a negative contribution to the spin polarization and thus will reduce the polarization spitting among $\L$, $\Xi^0$, and $\O^-$ or even violate the polarization ordering as shown in \fig{globpol} (right) which does not contain any magnetic field contribution.

\begin{figure}[h]
    \begin{center}
    %\hspace{-1.5cm}
    \includegraphics[width=4.cm]{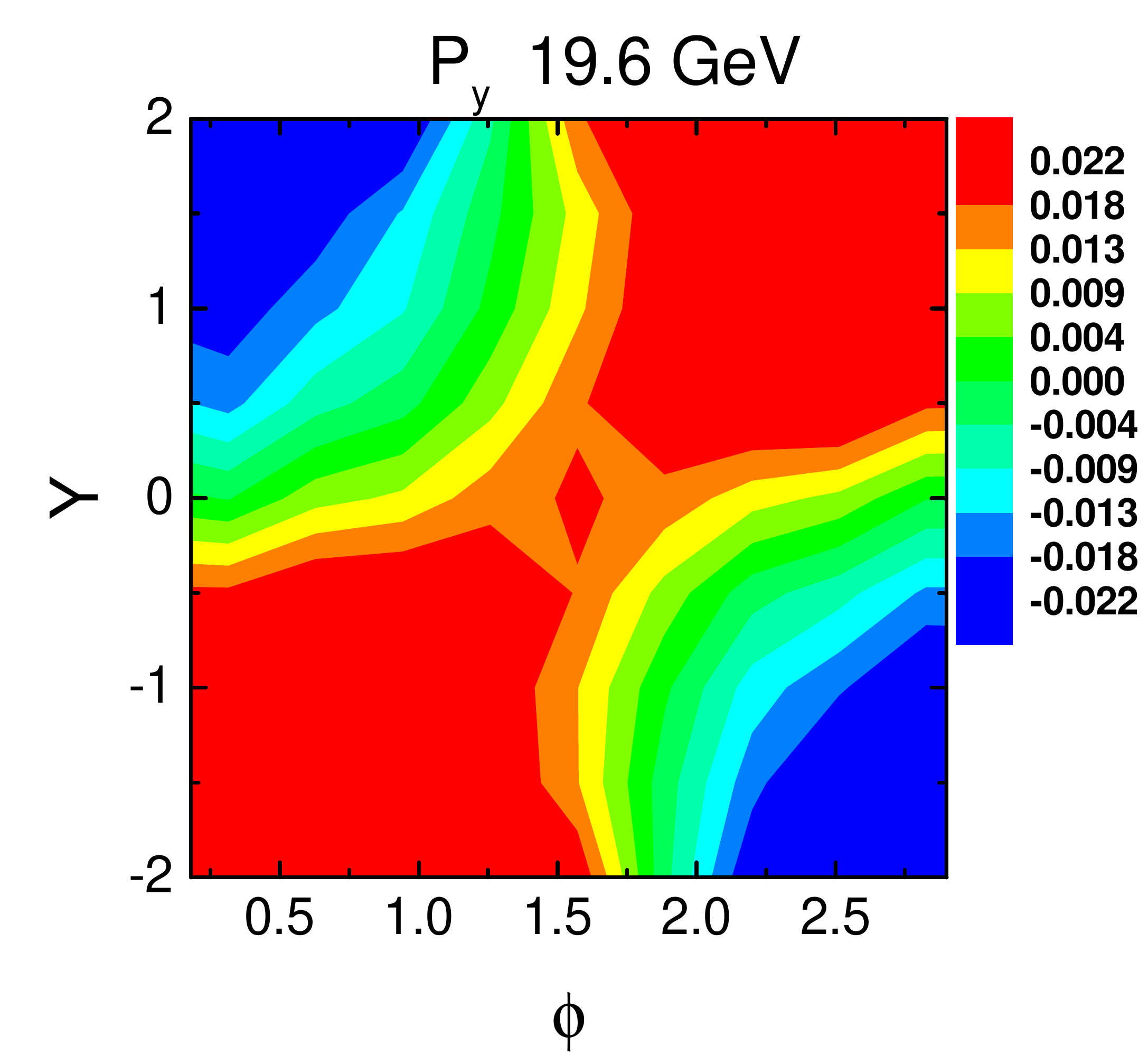} %\\
    \includegraphics[width=4.cm]{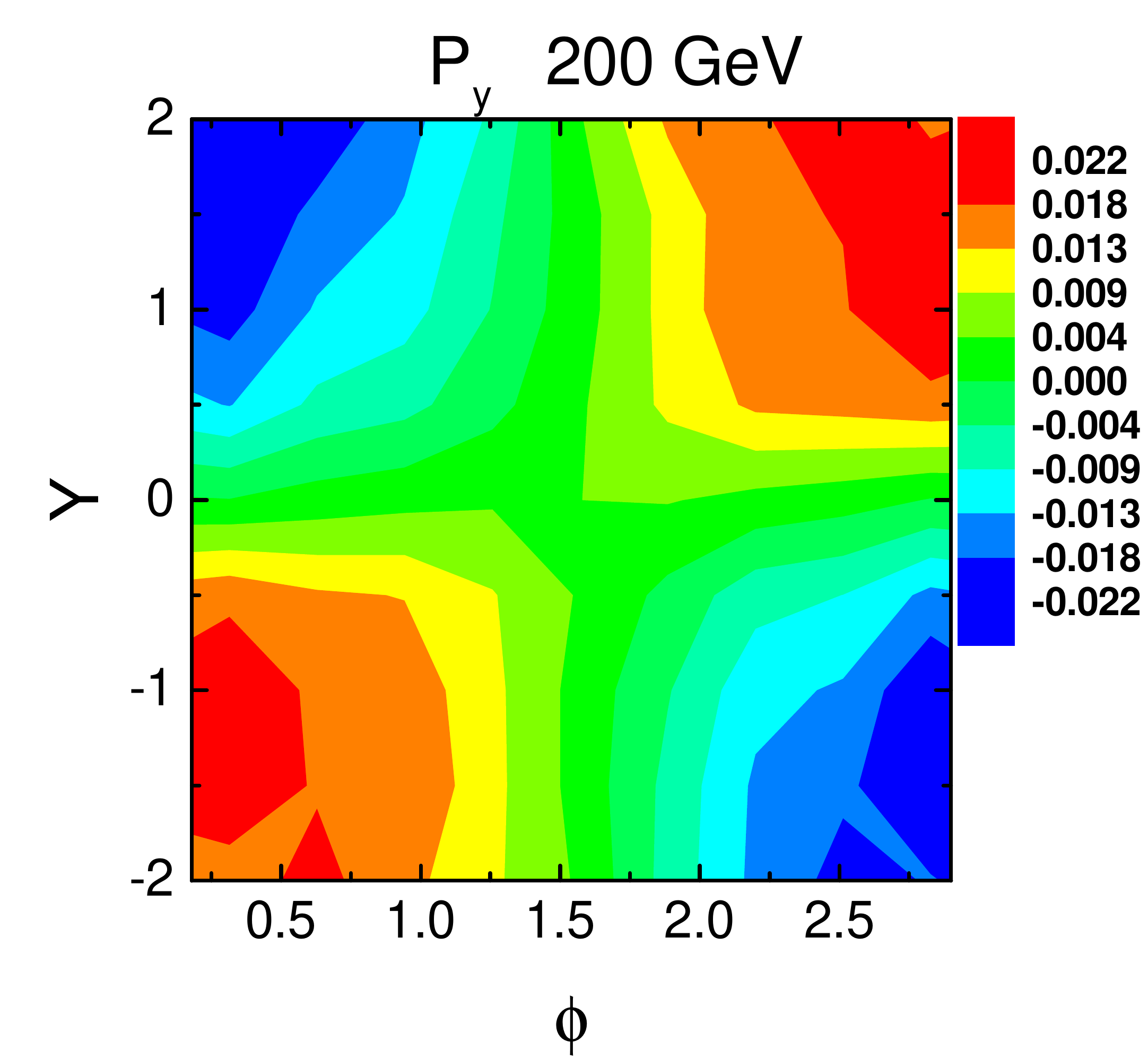}
    \caption{(Color online) The rapidity-azimuth distribution of the event-averaged spin polarization of $\Lambda$ and $\bar{\L}$ for Au + Au collisions at 20-50\% centrality range at 19.6 and 200 GeV, respectively. }
    \label{dis:pol}
    \end{center}
\end{figure}

\begin{figure}[h]
    \begin{center}
     \includegraphics[width=4cm]{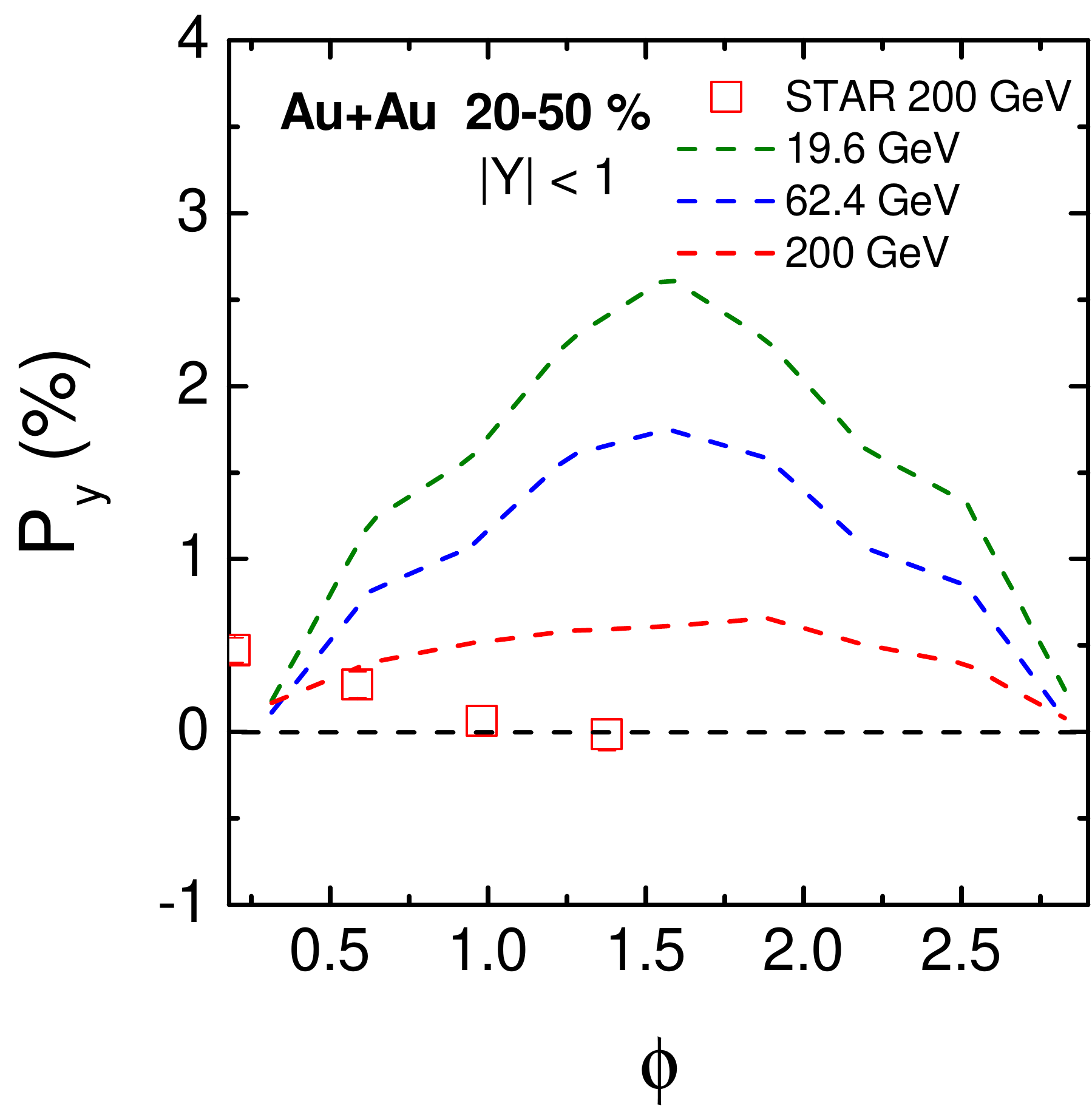}
    \caption{(Color online) The azimuthal angle dependence of $\L$ and $\bar{\L}$ polarization in rapidity region $|Y|<1$ for Au + Au collisions at 19.6, 62.4, and 200 GeV. The experimental data~\cite{Adam:2018ivw} is also shown.}
    \label{dis:pol2}
    \end{center}
    \end{figure}

Next, we study the final-state $\L$ and $\bar{\L}$ spin response to the vortical quadrupole in the partonic phase as shown in \fig{vorqua}. In \fig{dis:pol}, we show the distribution of event-averaged $P_y$ for $\L$ and $\bar{\L}$ in the rapidity-azimuth ($Y$-$\f$) plane for Au + Au collisions at $19.6$ and $200$ GeV and centrality $20$-$50$\%. Corresponding to \fig{vorqua} in coordinate space, the quadrupole in $P_y$ in momentum space is also clearly seen in \fig{dis:pol}. If we focus on the mid-rapidity region, e.g., $|Y|<1$, where the global OAM contribution could dominate, we find that $P_y$ increases from the in-plane direction to the out-of-plane direction, as shown in \fig{dis:pol2} which is, however, opposite to the experimental data. We note that similar opposite-to-experiment behavior of $P_y$ was also seen in the hydrodynamic simulations~\cite{Xie:2016fjj,Li:2017dan}. This discrepancy between theoretical calculations and experimental data is very puzzling. One issue that may affect the azimuthal dependence is that the spin polarization along the out-of-plane direction may be quenched by the hot medium which is not taken into account in the theoretical calculations. We will in future works study this puzzle.

Next, let us focus on the question: What is the specific hadronic observable for the smoke-loop vortical structure at finite spacetime rapidity? The rapidity odd feature of such a vortical structure suggests that the polarization weighted by the rapidity will be a good observable; such an idea was examined in Ref.~\cite{Xia:2018tes} and indeed, they found that the rapidity-sign weighted polarization is very large and has mild collision energy dependence. We here propose another observable for the smoke-loop type vortical structure, that is the spin harmonic coefficients at finite rapidity.

\begin{figure}[h]
    \begin{center}
%    \hspace{-1.5cm}
    \includegraphics[width=4cm]{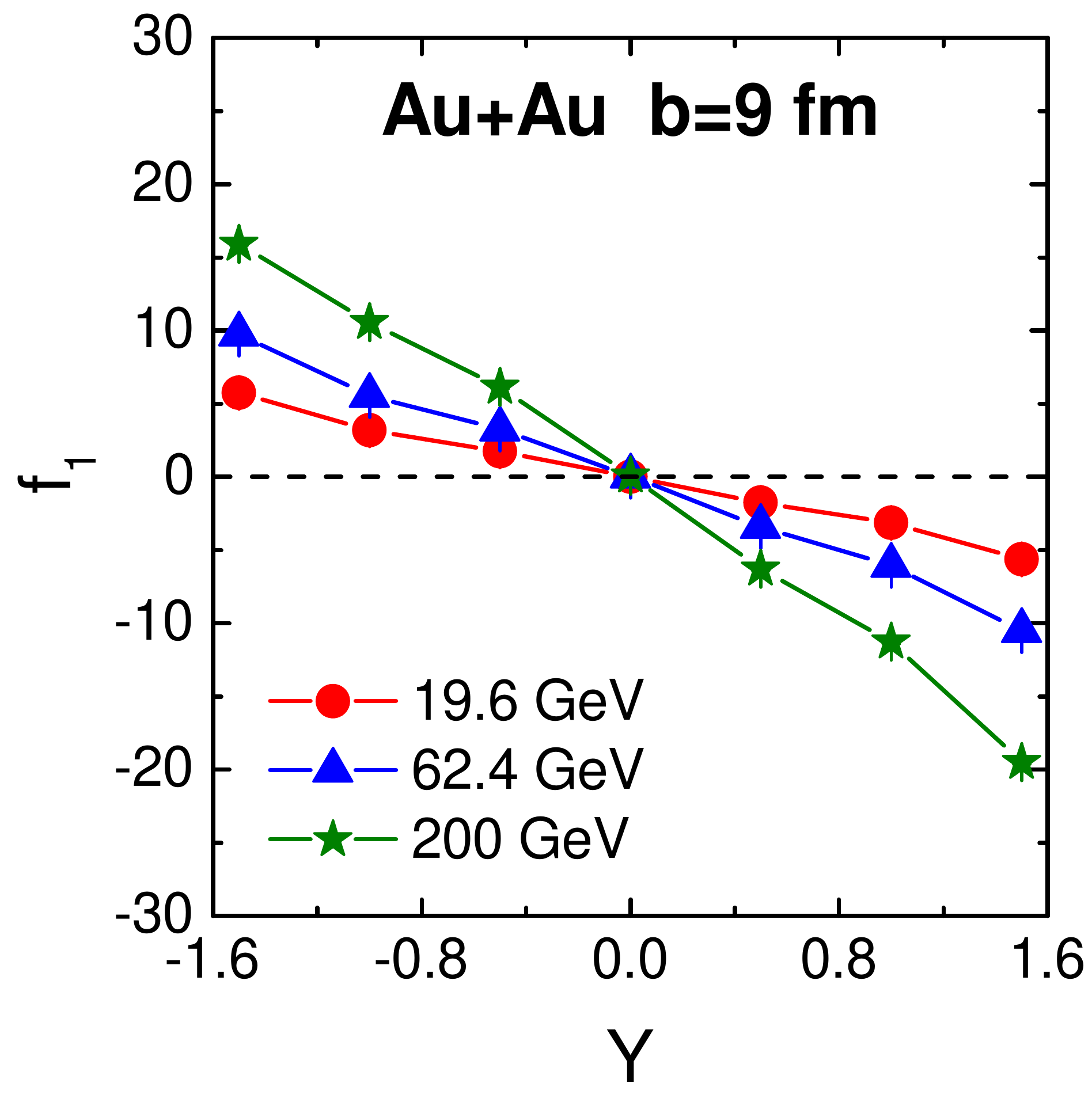} %\\
    \includegraphics[width=4cm]{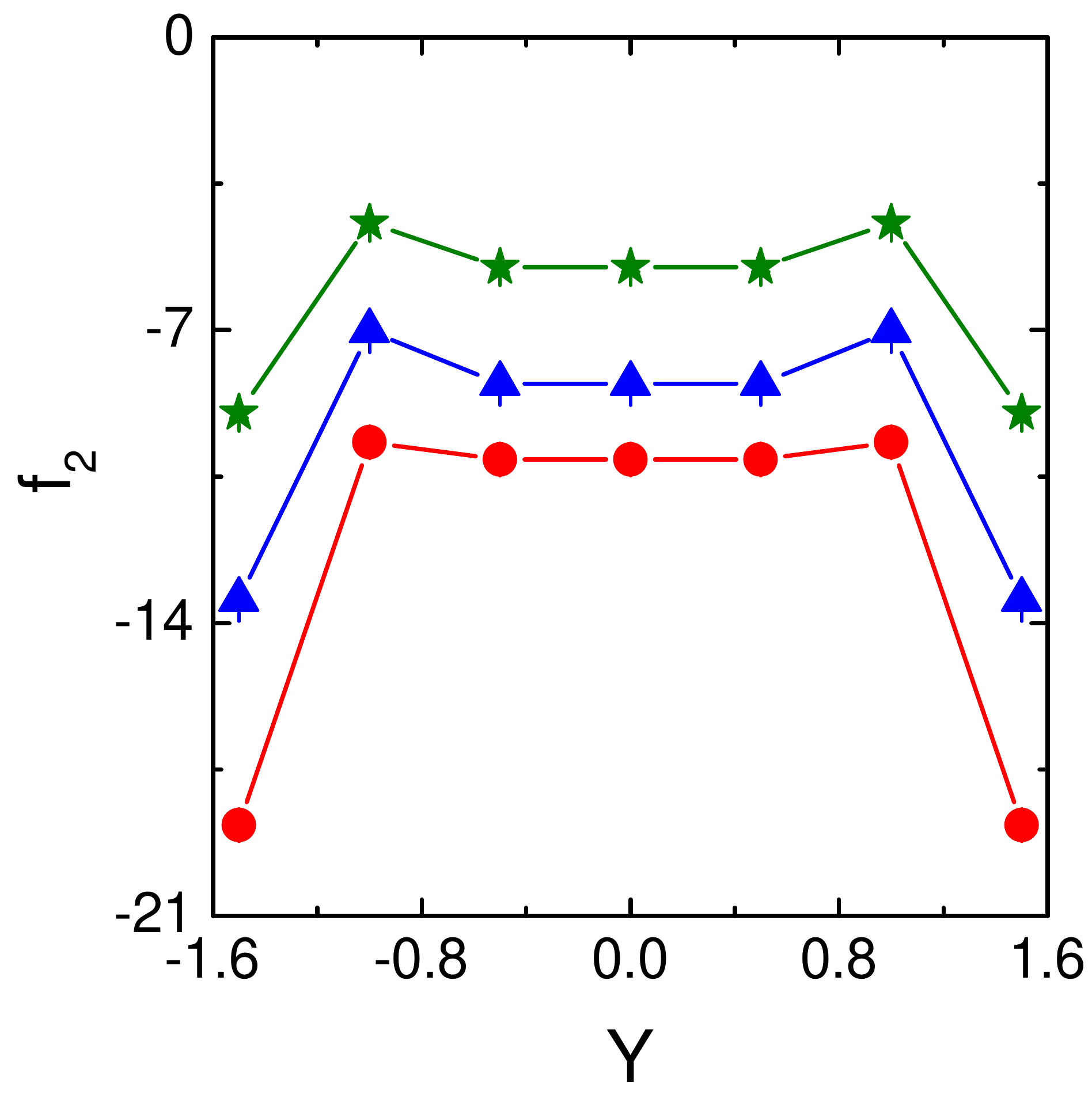} %\\
        \caption{(Color online) The directed and elliptic spin harmonic coefficients, $f_1$ and $f_2$, versus rapidity for Au + Au collisions with fixed impact parameter $b=9$ fm for $\sqrt{s}$ from 19.6 - 200 GeV.}
    \label{spin:har}
    \end{center}
\end{figure}

Recall that the charged particle distribution can be decomposed into different harmonic components as in \eq{vn:char} in which the harmonic coefficients reflect the response of the final-state momentum-space distribution to the initial anisotropy in coordinate space. Similarly, we can expect that the anisotropy in the vortical structure of the early or intermediate stage fluid can be reflected in the harmonic coefficients of the spin-polarization observable as given in
\begin{eqnarray}\label{Eqnarry35}
     P_{y}(Y,\f)= \frac{1}{2\pi}P_{y}(Y)\{1+2\sum_{n=1}^{\infty}f_{n}\cos[n(\f-\Phi_{n})]\},
\end{eqnarray}
where $\Phi_n$ defines the $n$th harmonic plane for spin and the corresponding harmonic coefficient is $f_n$. In real experiments and also in numerical simulations, the harmonic plane $\Phi_n$ would suffer from strong fluctuation as the numbers of $\L$ and $\bar{\L}$ (or other hadrons whose spin polarization can be measured) are small. Thus in the following simulation we will use $\Psi_n$ as defined in \eq{vn:char} to replace $\Phi_n$. In other words, we will study the harmonic flows of spin with respect to the harmonic plane determined by the distribution of charged hadrons. Thus we will calculate $f_n$ by using
\begin{eqnarray}\label{har:fn}
     f_n (Y) = \frac{\int d\f\cos[n(\f-\Psi_{n})]P_{y}(Y,\f)}{\int d\f P_y(Y,\f)}.
\end{eqnarray}
The results for the first two harmonics, $f_1$ and $f_2$, are shown in \fig{spin:har}. The directed flow of spin, $f_1$, which is induced by the vorticity owning to collective expansion, is odd in rapidity and peaks at finite rapidity in accordance with \fig{dis:pol}. It is sensitive to the collision energy as the azimuthal distribution at finite rapidity, as shown in \fig{dis:pol}, is. The measurement of the slope of $f_1(Y)$ versus rapidity at $Y=0$ may provide further constraint to the equation of state of the hot medium, especially the vortical susceptibility of the hot medium~\cite{Aristova:2016wxe}. The elliptic flow of spin, $f_2$, is even in rapidity. It is negative, in consistence with our numerical result in \fig{dis:pol2}; However, one should be noticed that the experimental data shows a opposite trend for the $\phi$ dependence of $P_y$ in mid-rapidity region which should result in a positive $f_2$. Again, this discrepancy will be examined in future works.

%It is interesting to notice that the rapidity dependence of the spin harmonic coefficients are very similar with the rapidity dependence of the harmonic flow coefficients $v_n$ of charged hadrons. Thus we make an appropriate ``ratio" of the two and define
%\begin{eqnarray}\label{EqnJ1}
%     J_{n}(Y) &=& \frac{\int d\f\langle \cos[n(\f-\Psi_{n})]P_{y}(Y,\f)\rangle}{\int d\f\langle \cos[n(\f-\Psi_{n})]\frac{d N_{\rm ch}}{d\f}(Y,\f)\rangle} .
%\end{eqnarray}
%The numerical results $J_{1,2}$ are shown in \fig{spin:har2}. It is seen that $J_1$ is now rapidity even and is less sensitive to the collision energy than $f_1$. This is consistent with the observation made in Ref.~\cite{Xia:2018tes} which shows that the rapidity-sign weighted polarization is insensitive to the collision energy. In the rapidity region $|Y|<1$, $J_2$ at higher collision energy is insensitive to rapidity in coincidence with the right panel of \fig{ptrap}. $J_2$ decreases with increasing collision energy near the mid-rapidity region, because the global OAM induced vorticity decrease with increasing $\sqrt{s}$.
%\begin{figure}[h]
%    \begin{center}
%    \hspace{-1.5cm}
%    \includegraphics[width=4.0cm]{fig111a} %\\
%    \hspace{0.0cm}
%    \includegraphics[width=4.0cm]{fig111b} %\\
%    \hspace{0.0cm}
%    \includegraphics[width=4.0cm]{fig111c} %\\
%    \hspace{0.0cm}
%    \includegraphics[width=4.0cm]{fig111d} %\\
%    \caption{(Color online) The $v_n$ corrected spin harmonics for Au + Au collisions with fixed impact parameter $b=9$ fm at $\sqrt{s}=$7.7 - 200 GeV.}
%    \label{spin:har2}
%    \end{center}
%\end{figure}

\section{Discussions} \label{sec:dis}
In this paper, we have systematically studied the event-by-event generation of the thermal vorticity in Au + Au collisions at different collisions energies. The thermal vorticity can have different sources among which the primary ones are the global OAM of the colliding system and the collective expansion of the fire ball. The former can give the global spin polarization of $\L$ and $\bar{\L}$ hyperons in the OAM direction in the mid-rapidity region and our numerical simulation can explain the experimental data quite well. The latter can lead to intriguing smoke-loop type vortical structure at finite spacetime rapidity which can drive a vortical quadrupole in the reaction plane. We propose to use the spin harmonic flows, especially the first and second order spin harmonics to detect such a quardrupolar vortical configuration.

However, it should be noted that there exist evident discrepancy between the theoretical results and the experimental data. For example, the azimuthal distribution of either the longitudinal spin polarization or the polarization along the OAM direction at the mid-rapidity region has opposite trend in theoretical results comparing to the recent experimental data~\cite{Adam:2018ivw}. Another example is that the spin-alignment measurement of the vector mesons $\f$ and $K^{*0}$ also show features that is in contradiction to the theoretical predictions~\cite{Liang:2004xn,Zhou:2017nwi,Singh:2018uad}. These puzzles indicate that our current understanding of the spin polarization mechanism and also the possible background effects may need careful reexamination. We will report our studies concerning these puzzles in the future.

{\bf Acknowledgments.---}  We thank F. Becattini, H. Li, A. Tang, Q. Wang, X.-L. Xia, and Z. Xu for useful discussions.
This work is supported by the Young 1000 Talents Program of China and by NSFC through Grants No. 11535012, No. 11675041, No. 11405066, and No. 11535005.

%=======================================reference=========================================

%=======================================================

\end{document}